\documentclass[epj]{svjour}
\usepackage{graphics}
\usepackage{euscript,amstext,amssymb,amsbsy}
\input{psfig}
\input{epsf}

\def\rmqp{{ \hbox{\rm q}'} }
\def\rmq{{\hbox{\rm q}}}
\def\qt0{\tilde{q}_0}

\def\calm{{\cal M}}
\def\dcalm{\Delta{\cal M}}
\def\Dirac#1{#1\hskip-5pt/}
\begin{document}
\title{Real and Virtual Compton Scattering off the Nucleon}
\author{Marc Vanderhaeghen
}                     
\institute{Institut f\"{u}r Kernphysik, Johannes Gutenberg-Universit\"{a}t, D-55099 Mainz, Germany}
\date{Received: July 19, 2000}
%
\abstract{A review is given of the very recent developments in the fields of
  real and virtual Compton scattering off the nucleon. 
Both real and virtual Compton scattering reactions are discussed at
  low outgoing photon energy 
where one accesses polarizabilities of the nucleon. 
The real Compton scattering at large momentum transfer 
is discussed which is asymptotically a tool to obtain 
information on the valence quark wave function of the nucleon. 
The rapid developments in deeply virtual Compton scattering and associated 
meson electroproduction reactions at high energy, high photon virtuality  
and small momentum transfer to the nucleon are discussed. 
A unified theoretical description of those
processes has emerged over the last few years, which gives access to new, 
generalized parton distributions. 
The experimental status and perspectives in these fields are also discussed. 
} 
\maketitle
\section{Introduction}
\label{introduction}

In the study of hadron structure, one of the main questions is 
how hadrons and nuclei are built from quarks and gluons 
and how one goes over from a description in terms of quark and gluon
degrees of freedom to a description in terms of hadronic degrees of 
freedom. 
\newline
\indent
Nowadays, precision experiments at high energy have established Quantum
Chromo Dynamics (QCD) as the gauge theory of strong interactions 
describing the dynamics between colored quarks and gluons. 
QCD exhibits the property that the interaction between the quarks
becomes weak at very short distances, which is known as 
asymptotic freedom, and which allows us to use perturbation theory to
describe high energy strong interaction phenomena. On the other hand 
at low energy, the QCD coupling constant grows, and quarks and gluons
are confined into colorless mesons and baryons, which are the
particles as seen through experiments. In this phase of hadronic
matter, the underlying chiral symmetry of QCD, due to the nearly massless
up, down and approximately also strange quarks, is spontaneously
broken. To describe this regime directly from the QCD Lagrangian is a
hard task which still defies a solution due to the strong coupling
constant requiring non-perturbative methods. Probably the most
promising and direct approach is the numerical solution of QCD
through lattice calculations. For static hadronic properties, such as
e.g. masses, much progress has already been made, but for more
complicated hadron structure quantities, such as e.g. nucleon parton
distributions, lattice calculations are still in their infancy. 
\newline
\indent
In absence of a full numerical solution of QCD, 
which would be able to describe the
rich complexity of the hadro\-nic many body systems 
from their underlying dynamics,
a complementary strategy to refine our understanding 
of hadron structure is to perform new types of
precision experiments in kinematical regimes at low energy, 
which require an inherent non-perturbative description. Besides, one may 
perform new types of experiments at high energies, 
in those kinematical regimes where 
factorization theorems have been proven. 
Such experiments will allow us to use an accurate perturbative QCD description 
of the reaction dynamics as a tool to 
extract new types of non-perturbative hadron structure information. 
\newline
\indent
In this quest at the intersection of particle and nuclear physics, 
the experiments performed with electromagnetic probes play an important role. 
A new generation of precision experiments 
has become possible with the advent of the new 
electron accelerators and in combination with 
high precision and large acceptance detectors.
In particular, high precision Compton scattering experiments 
have become a reality in recent years. 
In Compton scattering, a real or virtual photon interacts with the nucleon 
and a real photon is emitted in the process. As this is a purely 
electromagnetic process, it yields small cross sections (compared to
hadronic reactions), but on the other hand constitutes a clean probe of 
hadron structure.  
\newline
\indent
In this paper, a review will be given of very recent developments 
in the field of real and virtual Compton scattering off the nucleon. 
I shall discuss real and virtual Compton scattering at the same
time and point out their complementarity and the differences in the
extracted nucleon structure information.  
The emphasis is on those kinematical regimes where a fruitful interpretation
is terms of nucleon structure observables has been shown to be possible.  
Virtual Compton scattering (VCS) off the nucleon has been reviewed
before in Ref.~\cite{Gui98}, which is referred to for most technical details. 
For the VCS part, the emphasis is
on the progress over the past two years in the light of the first high
precision VCS data in the threshold regime now available, and
on the rapid development in the field of deeply virtual Compton
scattering (DVCS) at large photon virtuality.
\newline
\indent
In section \ref{rcsdr}, the real Compton scattering (RCS) 
process below pion threshold is discussed. In this regime, the RCS process 
can be interpreted as the global response of the nucleon to an applied
electromagnetic field, which allows us to access global nucleon 
polarizabilities. A dispersion relation formalism is described,
which was developed as a method to minimize the model uncertainty in
the extraction of nucleon polarizabilites 
from both unpolarized and polarized RCS data at low energy.
\newline
\indent
In section \ref{vcspol}, 
the virtual Compton scattering (VCS) reaction  
at low energy is discussed. It consists of a generalization 
of RCS in which both energy and momentum of the virtual photon 
can be varied independently, which
allows us to extract response functions, parametrized by 
the so-called generalized polarizabilities (GP's) of the nucleon. 
A first dedicated VCS experiment was performed at the MAMI accelerator,
and two combinations of those GP's have been measured. 
Their values are compared with nucleon structure model
predictions. Further experimental programs are underway 
at the major electron accelerators to measure the VCS observables. 
It is outlined how results of such experiments can be interpreted
in terms of the nucleon GP's, and in particular how polarization 
observables can separate the different GP's.
\newline
\indent
Besides the low energy region, RCS at high energy and large momentum transfer 
is a tool to access information on the partonic structure of the
nucleon. In section \ref{larget}, 
a leading order perturbative QCD calculation of RCS is 
described, which was developed to extract the valence 
quark wave function of the nucleon. Such an approach is compared with
competing mechanisms, and it is pointed out how the planned 
experiments can shed light on this field.
\newline
\indent
Section \ref{dvcs} discusses the recent developments in 
deeply virtual Compton scattering (DVCS) and associated meson
electroproduction reactions at high energy, high photon virtuality $Q^2$ 
and small momentum transfer to the nucleon. 
It is shown how a unified theoretical description of those
processes has recently emerged and how it gives access to new
parton distributions, the so-called skewed parton distributions, 
which are generalizations of the usual parton 
distributions as known from inclusive deep inelastic lepton scattering. 
Leading order perturbative QCD calculations of DVCS 
and different meson electroproduction reactions, 
using an ansatz for the skewed parton distributions, 
are discussed in the kinematical regimes accessible at present
or planned facilities. The corrections to those leading order QCD  
amplitudes and further open questions in this field are touched on briefly.   
\newline
\indent
In section \ref{radcorr}, 
the calculation of the QED radiative corrections to the VCS process is 
discussed, which is indispensable to accurately extract the nucleon
structure information from VCS experiments.
\newline
\indent
Finally, the conclusions and perspectives are given in section 
\ref{conclusion}.

\section{Real Compton scattering (RCS) and nucleon polarizabilities}
\label{rcsdr}

\subsection{Introduction}

In the study of nucleon structure, 
real Compton scattering (RCS) at low energy is a precision tool to
access global information on the nucleon ground state and its
excitation spectrum. RCS off the nucleon is determined by 6 independent
helicity amplitudes, which are functions of two variables, e.g.
the Lorentz invariant variables $\nu$ (related to the $lab$ energy of
the incident photon) and $t$ (related to the momentum transfer to the
target). In the limit $\nu\rightarrow 0$, corresponding to 
wavelengths much larger than the nucleon size, the general structure of
these amplitudes is governed by low energy theorems (LET) based on
Lorentz invariance, gauge invariance and crossing symmetry.
These theorems predict that the
leading terms of an expansion in $\nu$ are determined by global
properties of the nucleon, i.e. its charge, mass and anomalous
magnetic moment. Furthermore, the internal structure shows up only at
relative order $\nu^2$ and can be parametrized in terms of the 
polarizabilities. In this way, there appear 6 polarizabilities for the
nucleon, the electric and magnetic (scalar) polarizabilities
$\alpha$ and $\beta$ respectively, familiar from classical
physics, and 4 spin (vector) polarizabilities
$\gamma_1$ to $\gamma_4$, originating from the spin 1/2 nature of the
nucleon. These polarizabilities describe the response
of the nucleon's charge, magnetization, and spin distributions   
to an external quasistatic electromagnetic field. 
As such the polarizabilities are fundamental 
structure constants of the composite system.
\newline
\indent
The differential cross section for RCS in the limit 
$\nu\rightarrow 0$ is given by the (model independent) Thomson term, as
a consequence of the LET.
The electric and magnetic polarizabilities then appear, 
in a low-energy expansion for RCS, 
as interference between the Thomson term and the
subleading terms, i.e. as contribution of $O(\nu^2)$ in the
differential cross section. In this way, $\alpha$ and $\beta$ can in principle
be separated by studying the RCS angular distributions. However, it has
never been possible to isolate this term and thus to determine the
polarizabilities in a model independent way. The obvious reason is
that, for sufficiently small energies, say $\nu\leq 40$~MeV, the
structure effects are extremely small and hence the statistical errors
for the polarizabilities large. 
Therefore, one has to go to larger energies, where the higher terms in
the expansion become increasingly important and where also the spin
polarizabilities come into play. 
To determine the nucleon polarizabilities from RCS observables, a
reliable estimate of higher terms in the energy is therefore of utmost
importance. To this end, actual experiments were usually analyzed in 
an unsubtracted dispersion relation formalism at fixed $t$ \cite{Lvo97}. 
Using such an analysis, the proton scalar polarizabilities were derived  
from Compton scattering data below pion threshold, 
with the results \cite{Mac95}~:
\begin{eqnarray}
\alpha \;&=&\; \left(12.1 \,\pm\,0.8 \,\pm\, 0.5 \right)\,
\times\,10^{-4}\,{\rm fm}^3 \;, \nonumber\\  
\beta \;&=&\; \left(2.1 \,\mp\,0.8 \,\mp\,0.5 \right)\,
\times\,10^{-4}\,{\rm fm}^3 \;.
\label{eq1.1}
\end{eqnarray}
Very recently, new RCS data on the proton below pion threshold 
have become available \cite{Olm00}. 
These data increase the available world data set substantially, and 
yield, in an unsubtracted DR formalism, the results~:  
\begin{eqnarray}
\alpha \;&=&\; \left(11.89 \,\pm\, 0.57 \right)\,
\times\,10^{-4}\,{\rm fm}^3 \;, \nonumber\\  
\beta \;&=&\; \left(1.17 \,\pm\,0.75 \right)\,
\times\,10^{-4}\,{\rm fm}^3 \;.
\label{eq1.1b}
\end{eqnarray}
\indent
The sum of the scalar polarizabilities, 
which appears in the forward spin averaged 
Compton amplitude, can be determined directly from the 
total photoabsorption cross section by
Baldin's sum rule~\cite{Bal60}, which yields a rather precise value~:
\begin{eqnarray}
\alpha+\beta & = & \left(14.2 \,\pm\, 0.5 \right) \,
\times\,10^{-4}\,{\rm fm}^3 \;,  
\label{eq1.5a} \\
& = & \left(13.69 \,\pm\, 0.14 \right)\,
\times\,10^{-4}\,{\rm fm}^3 \;,
\label{eq1.5b}
\end{eqnarray}
with (\ref{eq1.5a}) from Ref.~\cite{Dam70} and 
(\ref{eq1.5b}) from Ref.~\cite{Bab98}.
\newline
\indent
Similarly, the proton forward spin polarizability can be evaluated by an
integral over the difference of the absorption cross sections in
states with helicity 3/2 and 1/2,
\begin{eqnarray}
\gamma_0 = \gamma_1-\gamma_2-2\gamma_4 
& = & -1.34 \, \times\,10^{-4}\,{\rm fm}^4 \;,
\label{eq1.6a} \\ 
& = & -0.80 \,\times\,10^{-4}\,{\rm fm}^4 \;,
\label{eq1.6b}
\end{eqnarray}
with (\ref{eq1.6a}) from Ref.~\cite{San94} and 
(\ref{eq1.6b}) from Ref.~\cite{Dre00a}.
While the predictions of Eqs.~(\ref{eq1.6a},\ref{eq1.6b}) 
rely on pion photoproduction multipoles, the
helicity cross sections have now been directly determined at MAMI by
scattering photons with circular polarizations on polarized 
protons~\cite{Are00}. The contribution to $\gamma_0$ for the proton 
within the measured integration range 
(200 MeV $\leq \nu \leq$ 800 MeV) is \cite{Are00}~: 
\begin{equation}
\gamma_0 \; {\big |}_{\, 200 \; {\rm MeV}}^{\, 800 \; {\rm MeV}} \;=\;
\left(-1.68 \pm 0.07 \right)\,\times\,10^{-4}\,{\rm fm}^4 \;.
\label{eq:go}
\end{equation}
The contribution below 200 MeV can be estimated with the  
HDT pion photoproduction multipoles \cite{Han98} to yield +1.0, 
and an estimate of 
the contribution above 800 MeV based on the SAID pion photoproduction
multipoles \cite{Arn96} yields -0.02, which results in a total value~: 
$\gamma_0$ = -0.7 (here and in the following, all spin
polarizabilities are given in units $10^{-4}\,{\rm fm}^4$).
\newline
\indent
Furthermore, unpolarized RCS data in the $\Delta$(1232)-region were
used to give - within a dispersion relation formalism - 
a first prediction for the so-called backward spin
polarizability of the proton, i.e. the particular combination
$\gamma_{\pi}=\gamma_1+\gamma_2+2\gamma_4$ entering the Compton
spin-flip amplitude at $\theta=180^{\circ}$~\cite{Ton98}~:
\begin{equation}
\gamma_{\pi} = - \left[ 27.1 \,\pm\, 2.2 ({\rm stat + syst}) 
{+2.8 \atop -2.4} ({\rm model})\right] 
\times\,10^{-4}\,{\rm fm}^4 .
\label{eq1.2}
\end{equation}
\indent
These values for the polarizabilities can be compared with our present day
theoretical understanding from chiral perturbation theory (ChPT). 
A calculation to $O(p^4)$ in heavy baryon ChPT (HBChPT),
where the expansion parameter $p$ is an external momentum or the quark
mass, yields (here and in the following, $\alpha$ and $\beta$
are given in units $10^{-4}\,{\rm fm}^3$)~: 
$\alpha=10.5\pm 2.0$ and $\beta=3.5\pm
3.6$, the errors being due to 4 counter terms entering to that order,
which were estimated by resonance saturation~\cite{Ber93}. 
One of these counter terms describes the large paramagnetic contribution of the
$\Delta$(1232) resonance, which is partly cancelled by large diamagnetic
contributions of pion-nucleon (N$\pi$)-loops.
In view of the importance of the $\Delta$ resonance, the calculation
was also done by including the $\Delta$
as a dynamical degree of freedom. This adds a further expansion
parameter, the difference of the $\Delta$ and nucleon masses
(``$\epsilon$ expansion''). A calculation to $O(\epsilon^3)$ yielded
$\alpha$ = 12.2 + 0 + 4.2 = 16.4 
and $\beta$ = 1.2 + 7.2 + 0.7 = 9.1, 
the 3 separate terms referring to contributions of N$\pi$-loops (which
is the $O(p^3)$ result), 
$\Delta$-pole terms, and $\Delta\pi$-loops~\cite{Hem97a,Hem98}.
These $O(\epsilon^3)$ predictions are
clearly at variance with the data, in particular $\alpha+\beta=25.5$
is nearly twice the rather precise value determined from Baldin's 
sum rule Eq.~(\ref{eq1.5b}). 
\newline
\indent
The spin polarizabilities have also been calculated in HBChPT. 
The $O(\epsilon^3)$ predictions for the proton are \cite{Hem98}~:
$\gamma_0  =  4.6-2.4-0.2+0=+2.0$, and
$\gamma_{\pi}  =  4.6+2.4-0.2-43.5=-36.7$,
the 4 separate contributions referring to
N$\pi$-loops ($O(p^3)$ result), 
$\Delta$-poles, $\Delta\pi$-loops, and the triangle
anomaly, in that order. It is obvious that the anomaly or $\pi^0$-pole
gives by far the most important contribution to $\gamma_{\pi}$, 
and that it would require surprisingly large higher order contributions 
to bring $\gamma_{\pi}$ close to the value of Eq.~(\ref{eq1.2}). 
Recently, the N$\pi$-loop contribution to the spin polarizabilities 
have been evaluated in HBChPT to $O(p^4)$ 
by several groups \cite{Ji00,Vij00,Gel00}. 
In Refs.~\cite{Ji00,Vij00}, the result for the
proton is $\gamma_0 = +4.5 - 8.4$, where the
two contributions are the $O(p^3)$ and $O(p^4)$ N$\pi$-loop
contributions, in this order.
Based on the large $O(p^4)$ correction term, the authors in 
\cite{Ji00,Vij00} call the convergence of the chiral expansion into question. 
However in Ref.~\cite{Gel00}, different results were obtained for the
$O(p^4)$ N$\pi$-loop contributions to the 4 spin polarizabilities. It
was argued that these differences 
are due to how one defines and extracts the $O(p^4)$ 
spin-dependent polarizabilities in chiral effective field theories. 
Following the procedure of Ref.~\cite{Gel00},
which removes first all one-particle reducible contributions from the
spin-dependent Compton amplitude, 
the resulting values for $\gamma_0$ and $\gamma_\pi$ of the proton are
$\gamma_0 = +4.6 - 5.6 = -1.0$, and 
$\gamma_\pi = +4.6 - 1.2 = +3.4$ (without the $\pi^0$-pole), 
the separate contributions being again the 
$O(p^3)$ and $O(p^4)$ N$\pi$-loop contributions respectively. 
For $\gamma_0$, a convergence of HBChPT at order $O(p^4)$ was not expected 
\cite{Gel00}, whereas the result for $\gamma_\pi$ - when adding the
$\pi^0$-pole contribution - is not compatible with the estimate 
of Eq.~(\ref{eq1.2}) obtained by Ref.~\cite{Ton98}.
\newline
\indent
In order to refine our present understanding of the nucleon
polarizabilities, a better understanding of the convergence 
of the HBChPT expansion is absolutely necessary, and it is to be hoped
that a calculation to $O(\epsilon^4)$ will clarify the status. 
On the other hand, it is also indispensable to minimize any model dependence
in the extraction of the polarizabilities from the data. To this end, 
a fixed-$t$ subtracted dispersion relation (DR) formalism 
was developed in Ref.~\cite{Dre00a} for RCS 
off the nucleon at photon energies below 500 MeV, as a
formalism to extract the nucleon polarizabilities with a minimum of
model dependence as is described in the following.

\subsection{Fixed-t subtracted dispersion relations for RCS}

To perform a dispersion theoretical analysis of Compton scattering, one
has to calculate the 6 independent structure functions $A_i(\nu,
t)$, $i=1,...,6$ (defined in Ref.~\cite{Lvo97}). 
They are functions of the usual Mandelstam variable $t$, and of $\nu$, 
defined in terms of the Mandelstam variables $s$ and $u$ as 
$\nu=(s-u)/(4m_N)$, with $m_N$ the nucleon mass. 
The invariant amplitudes $A_i$ are free of kinematical singularities
and constraints, and because of the crossing symmetry they satisfy the
relation $A_i(\nu, t)=A_i(-\nu, t)$. Assuming further analyticity and
an appropriate high-energy behavior, the amplitudes $A_i$ fulfill 
unsubtracted DR at fixed $t$~:
\begin{equation}
{\mathrm Re} A_i(\nu, t) \,=\, A_i^B(\nu, t) \,+\,
{2 \over \pi} \; {\mathcal P} \int_{\nu_{thr}}^{+ \infty} d\nu' \; 
{{\nu' \; {\mathrm Im}_s A_i(\nu',t)} \over {\nu'^2 - \nu^2}} \, ,
\label{eq:unsub} 
\end{equation}
where $A_i^B$ are the Born (nucleon pole) contributions, 
and where ${\mathrm Im}_s A_i$ are the discontinuities across the 
$s$-channel cuts of the Compton process, starting from 
the threshold for pion production, $\nu_{thr}$. 
However, such unsubtracted DR require that at high energies 
($\nu \rightarrow \infty$) the amplitudes ${\mathrm Im}_s A_i(\nu,t)$ 
drop fast enough such that the integral of Eq.~(\ref{eq:unsub}) is convergent 
and the contribution from the semi-circle at infinity can be neglected. 
For real Compton scattering, Regge theory predicts the following high-energy 
behavior for $\nu \rightarrow \infty$ and fixed $t$~\cite{Lvo97}~:
\begin{equation}
A_{1,2} \sim \nu^{\alpha(t)} \;,\hspace{.5cm}
A_{3,5,6} \sim \nu^{\alpha(t) - 2} \;,\hspace{.5cm} 
A_{4} \sim \nu^{\alpha(t) - 3} \;,
\label{eq:reggebehav}
\end{equation}
where $\alpha (t) \lesssim 1$ is the Regge trajectory. 
Due to the high-energy behavior of Eq.~(\ref{eq:reggebehav}), 
the unsubtracted dispersion integral of
Eq.~(\ref{eq:unsub}) diverges for the amplitudes $A_1$ and $A_2$. In order
to obtain useful results for these two amplitudes, L'vov et
al.~\cite{Lvo97} proposed to close the contour of the integral in
Eq.~(\ref{eq:unsub}) by a semi-circle of finite radius $\nu_{max}$
in the complex plane (instead of the usually assumed infinite radius!),  
i.e.  the real parts of $A_1$ and $A_2$ are calculated from the decomposition
\begin{equation}
{\mathrm Re} A_i(\nu, t) \;=\; A_i^B(\nu, t) \;+\;
A_i^{int}(\nu, t) \;+\; A_i^{as}(\nu, t) \;,
\label{eq:aintas}
\end{equation}
with $A_i^{int}$ the $s$-channel integral from pion
threshold $\nu_{thr}$ to a finite upper limit $\nu_{max}$,
and an `asymptotic contribution' $A_i^{as}$ representing the
contribution along the finite semi-circle of radius $\nu_{max}$ in the
complex plane. In the actual calculations, the $s$-channel integral is
typically evaluated up to a maximum photon energy of about  
$1.5$~GeV, for which the imaginary part
of the amplitudes can be expressed through unitarity by meson
photoproduction amplitudes (mainly 1$\pi$ and 2$\pi$ photoproduction)
taken from experiment.  All contributions from higher energies are
then absorbed in the asymptotic terms $A_i^{as}$, 
which are replaced by a finite 
number of energy independent poles in the $t$ channel. In particular
the asymptotic part of $A_1$ is parametrized 
by the exchange of a scalar particle in the $t$ channel, i.e. an effective
``$\sigma$ meson''~\cite{Lvo97}. 
In a similar way, the asymptotic part of $A_2$ is described  by the $\pi^0$ 
$t$-channel pole. 
This procedure is relatively safe for $A_2$ because of the dominance
of the $\pi^0$ pole or triangle anomaly, which is well established
both experimentally and on general grounds as Wess-Zumino-Witten term.
However, it introduces a considerable model-dependence in the case of
$A_1$.
\newline
\indent
It was therefore the aim of Ref.~\cite{Dre00a} to avoid the
convergence problem of unsubtracted DR and the
phenomenology necessary to determine the asymptotic contribution. 
To this end, it was proposed to consider
DR's at fixed $t$ that are once subtracted at $\nu=0$,
\begin{eqnarray}
{\mathrm Re} A_i(\nu, t) \;&=&\; A_i^B(\nu, t) \;+\;
\left[ A_i(0, t) - A_i^B(0, t) \right] \nonumber\\
&+&\,{2 \over \pi} \;\nu^2\; 
{\mathcal P} \int_{\nu_{thr}}^{+ \infty} d\nu' \; 
{{\; {\mathrm Im}_s A_i(\nu',t)} \over {\nu' \; (\nu'^2 - \nu^2)}} \, .
\label{eq:sub} 
\end{eqnarray}
These subtracted DR should converge for all 6 invariant amplitudes
due to the two additional powers of $\nu'$ in the denominator, and they are
essentially saturated by the $\pi N$ intermediate states. 
In other words, the lesser known contributions of two and more pions 
as well as higher continua are small and may be treated reliably
by simple models.
\newline
\indent
The price to pay for this alternative is the appearance of the
subtraction functions $A_i(\nu=0, t)$, which have to be determined at
some small (negative) value of $t$. This was achieved  
by setting up once-subtracted DR, this time in the variable $t$ \cite{Dre00a}~:
\begin{eqnarray}
&&A_i(0, t) - A_i^B(0, t) = 
a_i \,+\, a_i^{t-pole} \nonumber\\
&&\;+\;{t \over \pi} \left( \int_{(2 m_\pi)^2}^{+ \infty} dt' 
- \int_{- \infty}^{-2 m_{\pi}^2 - 4 M m_\pi} dt' \right) 
{{{\mathrm Im}_t A_i(0,t')} \over {t' \, (t' - t)}} , \nonumber\\
&&
\label{eq:subt} 
\end{eqnarray}
where the six coefficients $a_i \equiv A_i(0, 0) - A_i^B(0, 0)$ are
simply related to the six polarizabilities $\alpha, \beta, \gamma_1,
\gamma_2, \gamma_3, \gamma_4$ (see Ref.~\cite{Dre00a} for details), and 
where $a_i^{t-pole}$ represents, in the case of $A_2$, 
the contribution of the $\pi^0$ pole in the $t$-channel.  
\newline
\indent
To evaluate the dispersion integrals, 
the imaginary part of the Compton amplitude 
due to the $s$-channel cuts in Eq.~(\ref{eq:sub})
is determined, through the unitarity relation, from the 
scattering amplitudes of photoproduction on the nucleon. 
Due to the energy denominator 
$1/\nu'(\nu'^2-\nu^2)$ in the subtracted dispersion integrals, 
the most important contribution is from the $\pi
N$ intermediate states, while mechanisms involving more pions or
heavier mesons in the intermediate states 
are largely suppressed.  
In Ref.~\cite{Dre00a}, the $\pi N$ contribution was then evaluated 
using the pion photoproduction multipole amplitudes of Ref.~\cite{Han98} 
at photon energies below 500 MeV, 
and at the higher energies using the SAID multipoles 
(SP98K solution)~\cite{Arn96} as input. 
The multipion channels (in particular the $\pi \pi N$ channels) 
were approximated by the inelastic
decay channels of the $\pi N$ resonances. It was found, however, that
in the subtracted DR formalism, the sensitivity to the multipion
channels is very small and that subtracted DR are essentially
saturated at $\nu \approx$ 0.4 GeV.   
\newline
\indent
The subtracted $t$-channel dispersion integral in
Eq.~(\ref{eq:subt}) from $4 m_\pi^2$ to $+ \infty$ is  
essentially saturated by the imaginary part of the $t$-channel amplitude 
$\gamma \gamma \rightarrow N \bar N$ due to 
$\pi\pi$ intermediate states. 
The dependence of the subtraction functions on momentum 
transfer $t$ can be calculated by including the experimental information 
on the $t$-channel process through $\pi \pi$ intermediate states as 
$\gamma \gamma \rightarrow \pi \pi \rightarrow N \bar N$. 
In Ref.~\cite{Dre00a}, a unitarized amplitude for the 
$\gamma \gamma \rightarrow \pi \pi$ subprocess was constructed, 
and a good description of the available data was found. 
This information is then combined with the 
$\pi \pi \rightarrow N \bar N$ amplitudes determined from dispersion
theory by analytical continuation of $\pi N$ scattering.  
In this way, one avoids the uncertainties in Compton scattering 
associated with the two-pion continuum in the $t$ channel, usually 
 modeled through the exchange of a somewhat fictitious $\sigma$ meson. 
The second integral in Eq.~(\ref{eq:subt}) extends from 
$- \infty$ to $-2\, (m_{\pi}^2 + 2 M m_\pi) \approx - 0.56$~GeV$^2$. 
As we address Compton scattering for photon energies below 
about 500 MeV, the value of $t$ stays sufficiently small 
so that the denominator in the integral provides a rather 
large suppression, resulting in a small contribution. The
contribution along the negative $t$-cut is estimated in 
the calculations \cite{Dre00b} by saturation with 
$\Delta$ intermediate states. 
Altogether the remaining uncertainties in 
the $s$- and $t$- channel subtracted integrals 
due to unknown high-energy contributions, are estimated to be less than 1\%. 
As a consequence, this subtracted DR formalism provides a direct 
cross check between Compton scattering and one-pion photoproduction. 
\newline
\indent
Although all 6 subtraction constants $a_1$ to $ a_6$ of
Eq.~(\ref{eq:subt}) could be used as fit parameters in the 
present formalism, the fit was restricted 
to the parameters $a_1$ and $a_2$, or equivalently to
$\alpha - \beta$ and $\gamma_\pi$ in \cite{Dre00a}.  
The subtraction constants $a_4,
a_5$ and $a_6$ were calculated through an unsubtracted
sum rule (Eq.~(\ref{eq:unsub}) for $\nu = t = 0$). 
The remaining subtraction constant $a_3$, 
related to $\alpha + \beta$ by  
$\alpha + \beta = - (a_3 + a_6)/(2 \pi)$, 
was fixed through Baldin's sum rule \cite{Bal60}, using the
value $\alpha + \beta = 13.69 \times 10^{-4}$ fm$^3$ \cite{Bab98}.

\subsection{Results for RCS observables}

\begin{figure}[ht]
\vspace{-.5cm}
\epsfysize=11 cm
\centerline{\hspace{.25cm} \epsffile{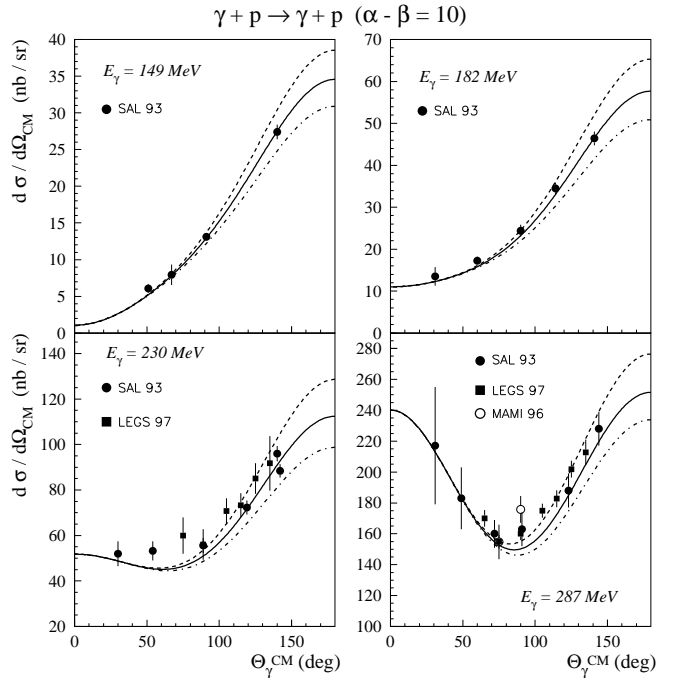}}
\vspace{-1.3cm}
\caption[]{Differential cross section for Compton scattering off the proton as 
function of the c.m. photon angle for different lab energies. 
The total results of the subtracted DR formalism 
are presented for fixed $\alpha - \beta = 10$   
and different values of $\gamma_\pi$ : 
$\gamma_\pi$ = -37 (dashed-dotted lines), 
$\gamma_\pi$ = -32 (full lines) 
and $\gamma_\pi$ = -27 (dashed lines). 
The references to the data can be found in \cite{Dre00a}.}
\label{fig:compton_delta_gpi}
\end{figure}

Since the polarizabilities enter as subtraction constants, the subtracted 
dispersion relations can be used to extract the nucleon 
polarizabilities from RCS data with a minimum of model dependence. 
The present formalism can be applied up to photon energies 
of about 500 MeV. 
\newline
\indent
Below pion production threshold, RCS data were usually analyzed to
extract $\alpha$ and $\beta$. However, it was shown that 
the sensitivity to $\gamma_\pi$ is not at all negligible,
especially at the backward angles and the higher energies, so that
both $\alpha - \beta$ and $\gamma_\pi$ should be fitted
simultaneously \cite{Dre00a}. 
\newline
\indent
RCS above pion threshold can serve as a complement to determine 
the polarizabilities, in particular the spin polarizabilities, 
and can provide valuable cross checks between Compton
scattering and pion photoproduction, provided 
one can minimize the model uncertainties in the dispersion formalism. 
The three types of dispersion integrals of Eqs.~(\ref{eq:sub}) 
and (\ref{eq:subt}) in the formalism outlined here 
are evaluated as described above. 
As a representative result obtained within the subtracted DR
formalism, the RCS differential cross sections above pion
threshold are shown in Fig.~\ref{fig:compton_delta_gpi} 
at fixed $\alpha - \beta$ = 10 (here and in the following in units of 
$10^{-4}$ fm$^3$), while $\gamma_\pi$ is varied between $-27$ 
(here and in the following in units of $10^{-4}$ fm$^4$) and $-37$ 
(for more details, see Refs.~\cite{Dre00a,Dre00b}). 
By comparing all available data above pion threshold, it was concluded
\cite{Dre00a} that there is no consistency between the pion
photoproduction data from MAMI (entering through the dispersion
integrals) and available Compton scattering data, 
in particular when comparing with the LEGS data, which
were used in the extraction of Eq.~(\ref{eq1.2}) for $\gamma_\pi$.  
Therefore, new data in the $\Delta$ region are called for,  
some of which have recently become available \cite{Wis99}. An 
analysis of those unpolarized data in a dispersion formalism favors 
also a much more negative value for $\gamma_\pi$
than extracted in Eq.~(\ref{eq1.2}). The fit performed by Ref.~\cite{Wis99}
yields $\alpha - \beta$ = 9.1 $\pm$ 1.7(stat + syst) 
$\pm$1.2(mod), when using a value of $\gamma_\pi = -37.6$.
 
Besides the existing information from unpolarized data, 
a full study of the spin (or vector) polarizabilities 
will however require double polarization experiments. 
It was in fact shown \cite{Dre00a} that 
the scattering of polarized photons on polarized protons 
is very sensitive to $\gamma_\pi$, in particular in the backward hemisphere 
and at energies between threshold and the $\Delta$ region.
In addition, possible normalization problems can be avoided by
measuring appropriate asymmetries.
Therefore, future polarization experiments hold the promise   
to disentangle the scalar and vector polarizabilities of the
nucleon with the help of the described subtracted DR formalism, 
and to further quantify the nucleon spin response 
in an external electromagnetic field.

\subsection{Higher order polarizabilities of the proton}

As outlined above, the electric and magnetic polarizabilities arise as 
$O(\nu^2)$ corrections to the lowest order (Thomson) scattering
amplitude. One can then ask the question whether Compton
scattering can also provide additional proton structure information
via the use of higher-order polarizabilities. 
If one extends the analysis to consider spin-averaged 
$O(\nu^4)$ terms (see Ref.~\cite{Hol00} for details), 
four new structures fulfill 
the requirements of gauge, P, and T invariance. 
Two new quantities, $\alpha_{E\nu}^p$ and 
$\beta_{M\nu}^p$, represent dispersive corrections to
the lowest order static polarizabilities, $\alpha$ and $\beta$ respectively,
and describe the response of the system to time-dependent fields.  
Two more quantities $\alpha_{E2}^p$, $\beta_{M2}^p$, represent 
quadrupole polarizabilities and measure the electric and magnetic quadrupole
moments induced in a system by the presence of an applied field
gradient. 
\newline
\indent
As to the experimental evaluation of such structure probes, it is, of
course, in principle possible to extract them directly from 
Compton cross section measurements. 
However, it is already clear from the previous discussion 
of present data, that isolating such pieces from other terms which 
affect the cross section at energies above $\sim$ 100 MeV is virtually 
impossible since additional higher order effects soon become equally 
important.
Thus an alternative procedure is required, which is made possible by
the validity of dispersion relations. Within the subtracted DR
formalism of Ref.~\cite{Dre00a} outlined above, those {\it higher order}
terms in the expansion of the Compton amplitudes $A_i$
can be reasonably evaluated as in \cite{Hol00}. 
These higher order polarizabilities can be expressed in
terms of appropriate derivatives of the RCS invariant amplitudes
$A_i$ at $t,\nu^2=0$, denoted by $a_{i,t}, a_{i,\nu}$. 
In terms of subtracted DR's, they take the form~:
 \begin{eqnarray}
\label{nu_int}
a_{i,\nu}&=&{2\over \pi}\int_{\nu_{thr}}^\infty d\nu'{{\rm
    Im}_sA_i(\nu',t=0)\over \nu'^3},\\
\label{t_int}
a_{i,t}&=& {1\over \pi}\left(\int_{4 m_\pi^2}^\infty
-\int_{-\infty}^{- 4 M m_\pi - 2 m_\pi^2} dt'
{{\rm Im}_tA_i(0,t')\over t'^2}\right) .  \;\;\;
\end{eqnarray}
The higher order polarizabilities are then obtained as 
linear combinations of the 
$a_{i,t}$ and $a_{i,\nu}$ (for details see \cite{Hol00}). 
The subtracted DR in Eqs.~(\ref{nu_int}, \ref{t_int}) were evaluated
as described above and yield (all in units of $10^{-4}$ fm$^5$)~:
\begin{eqnarray}
\label{DR_quadr}
\quad\alpha_{E\nu}^p &=& \,-3.84-0.19+0.06, \nonumber\\
\quad\beta_{M\nu}^p &=& +9.29+0.15-0.07, \nonumber\\
\alpha_{E2}^p &=& +29.31-0.10-0.17, \nonumber\\ 
\beta_{M2}^p &=& -24.33+0.10-0.34\, , 
\end{eqnarray}
where the second and third entries on the {\it rhs} of
Eq.~(\ref{DR_quadr}) estimate the uncertainties in the $s$- and
$t$-channel dispersion integrals. 
\newline
\indent
The values of Eq.~(\ref{DR_quadr}) were then confronted in \cite{Hol00} 
with the predictions of HBChPT at ${\cal O}(p^3)$
\begin{eqnarray}
{\cal O}(p^3): &&\quad\alpha_{E\nu}^p=2.4,\quad\beta_{M\nu}^p=3.7, \nonumber\\
&&\quad\alpha_{E2}^p=22.1,\quad\beta_{M2}^p=-9.5 \, .
\label{eq:quadpol_chpt}
\end{eqnarray}
By comparing Eq.~(\ref{DR_quadr}) and (\ref{eq:quadpol_chpt}) one finds 
that the size of $\alpha_{E2}^p$ is about right, while for
both $\beta_{M2}^p$ and $\beta_{M\nu}^p$ the sign and order of magnitude
is correct but additional contributions are called for.  The most
serious problem lies in the experimental determination of
$\alpha_{E\nu}$ which is negative in contradistinction to the chiral
prediction and to sum rule arguments which assert its positivity. 
To see if inclusion of $\Delta(1232)$ degrees of freedom 
can help to resolve the above discrepancies, these quantities
were also calculated in \cite{Hol00} in HBChPT to $O(\epsilon^3)$, 
with the result~:
\begin{eqnarray}
{\cal O}(\epsilon^3):
&&\quad \alpha_{E\nu}^p=1.7,\quad\beta_{M\nu}^p=7.5, \nonumber\\
&&\quad \alpha_{E2}^p=26.2,\quad\beta_{M2}^p=-12.3 \, .
\end{eqnarray}
Except for the sign problem with $\alpha_{E\nu}^p$ indicated above, 
which persists in the 
$\epsilon$-expansion, the changes are generally helpful, although the magnetic 
quadrupole polarizability is still somewhat underpredicted.
\newline
\indent
In \cite{Hol00}, the described analysis was also extended to  
higher order contributions $O(\nu^5)$ to the proton spin
polarizabilities, for which 8 new structures were found. 
A dispersive evaluation of those higher order spin polarizabilities
showed a qualitative agreement with HBChPT $O(\epsilon^3)$ predictions. 
\newline
\indent
Recently, an evaluation of the higher order polarizabilities 
of the proton in HBChPT to $O(p^4)$ has been reported \cite{Hem00b}, 
providing an important new testing ground for the chiral predictions.
It was found \cite{Hem00b} that the $O(p^4)$ HBChPT result for the 4
quadrupole polarizabilities and the 8 spin 
polarizabilities at $O(\nu^5)$ of the proton are in encouraging 
good agreement with the DR estimates of Ref.~\cite{Hol00}.
\newline
\indent
In summary, the subtracted DR formalism presented not only
provides a formalism to extract the lowest order nucleon
polarizabilities from present and forthcoming RCS data with a minimum
of model dependence. It can also be used to obtain information about
higher order polarizabilities of the proton, in this way providing a
great deal of additional nucleon structure information.  

\newpage

\section{Virtual Compton scattering (VCS) and generalized nucleon polarizabilities}
\label{vcspol}

\subsection{Introduction}

The nucleon structure information obtained through RCS, 
as discussed in section \ref{rcsdr}, can be generalized by  
virtual Compton scattering (VCS) below pion threshold.
VCS can be interpreted as electron scattering off a target polarized
by the presence of constant electric and magnetic fields. 
To see how VCS generalizes the RCS process, it is useful to think of
the analogy with the electromagnetic form factors. 
Their measurement through elastic electron-nucleon scattering 
reveals the spatial distribution of the charge and magnetization 
distributions of the target, whereas a real photon is only
sensitive to the overall charge and magnetization of the target. 
The physics addressed with VCS is then the same as if one were 
performing an elastic electron scattering experiment on a target 
placed between the plates of a capacitor or between the poles of a magnet.
In this way one studies the 
spatial distributions of the polarization densities of the
target, by means of the generalized polarizabilities, which are functions 
of the square of the four-momentum, $Q^2$, transferred by the electron. 
\newline
\indent
Experimentally, the VCS process is accessed through the 
electroproduction of photons, and we
consider in all of the following the reaction on a proton target,
i.e. the reaction $e p \to e p \gamma$. 
One immediately sees a difference with regard to 
the RCS $\gamma p \to \gamma p$ reaction, 
because in the $e p \rightarrow e p \gamma$ reaction, 
the final photon can be emitted either by the proton, giving access
to the VCS process, or by the electrons, which is referred to as the
Bethe-Heitler (BH) process. The BH amplitude can be calculated exactly
in QED, provided one knows the elastic form factors of the proton. Therefore
it contains no new information on the structure. Unfortunately, light
particles such as electrons radiate much more than the heavy proton.
Therefore the BH process generally dominates or 
at least interferes strongly with the
VCS process, and this may complicate the interpretation of the 
$e p \rightarrow e p \gamma$ reaction. The only way out of
this problem is either to find kinematical regions where the BH process
is suppressed or to have a very good theoretical control over the
interference between the BH and the VCS amplitudes, as will be
discussed below. 
\newline
\indent
Assuming that this problem is fixed, one can then proceed to extract  
the nucleon structure information from VCS. 
In doing so, care has to be taken to separate the
trivial response of the target, due to its  
global charge and/or a global magnetic moment. 
Indeed, if we put a proton in an electric field, 
the first effect we observe is that it moves as a whole.
Similarly, the magnetic field produces a precession of the magnetic
moment. This problem is absent when one studies 
the polarizability of a macroscopic sample because it can be
fixed in space by appropriate means, which is not possible for the proton.
This absence of a restoring force explains why the trivial response
due to the motion of charge and magnetic moment dominates
over the response of the internal degrees of freedom. This is the
physical origin of the low energy theorem (LET) \cite{Low58} for VCS. 
All what is needed to calculate this part of the response, 
are the parameters which control the motion, that is
the mass, the charge, and the magnetic moment. Once the motion is
known, one can compute the amplitude for scattering an electron
on this moving proton, the so-called Born amplitude. 
Having separated the trivial response, one can
parametrize the structure part of interest in the VCS
process through the so-called generalized polarizabilities (GP's) 
as in Ref.~\cite{Gui95}. 

\subsection{Definition of generalized polarizabilities}

The known BH + Born parts of the VCS amplitude at low energy 
start at order 1/$\rmqp$ in an expansion 
in the outgoing photon energy $\rmqp$. 
The LET \cite{Low58} asserts that the non trivial part of the VCS
amplitude, the so-called non-Born part (denoted by $H_{NB}$), begins at
order $\rmqp$. There is of course also a contribution
of order \( \rmqp \) in the BH + Born amplitude, but this term 
is exactly known and therefore can be subtracted, at least in principle. 
So what is needed next is an adequate parametrization
of \( H_{NB} \). 
For this purpose, a multipole expansion (in the c.m. frame) was performed in
\cite{Gui95} in order to 
take advantage of angular momentum and parity conservation. 
The behaviour of the non-Born VCS amplitude $H_{NB}$ 
at low energy ($\rmqp \to 0$) but at arbitrary three-momentum \( \rmq \) 
of the virtual photon, was then parametrized by 
10 functions of \( \rmq  \) defined by~:
\begin{equation}
\label{eq_3_35}
\hbox {\rm Limit\, of}\, \, \, \, \frac{1}{\rmqp }\frac{1}{\rmq
^{L}}H_{NB}^{(\rho' 1,\rho L)S}(\rmqp ,\, \rmq )\, 
\, \, \, \hbox {\rm when}\, \, \rmqp \rightarrow 0.
\end{equation}
In this notation, $\rho$ ($\rho'$) refers to the
electric (2), magnetic (1) or longitudinal (0) nature of the initial 
(final) photon, $L$ ($L' = 1$) represents the angular momentum of the
initial (final) photon, whereas $S$ differentiates between the 
spin-flip ($S=1$) and non spin-flip ($S=0$) 
character of the electromagnetic transition at the nucleon side. 
As the angular momentum of the outgoing photon is $L'$ = 1, 
this leads to 10 $\rmq$-dependent GP's, denoted generically 
by $P^{(\rho' \, L', \rho \,L)S}(\rmq)$. 
By imposing the constraints due to nucleon crossing
combined with charge conjugation invariance on the VCS
amplitude, it was shown however in \cite{Dre97,Dre98b} that 
4 of the GP's can be eliminated. Thus only 6 GP's , e.g. \cite{Gui98}
\begin{eqnarray}
&&P^{(01,01)0}(\rmq),\ P^{(11,11)0}(\rmq), \nonumber\\
&&P^{(01,01)1}(\rmq),\ P^{(11,11)1}(\rmq),\ 
P^{(11,02)1}(\rmq),\ P^{(01,12)1}(\rmq), \;\;\;\;\;\;\;\;
\label{eq_3_41}
\end{eqnarray}
are necessary to give the low energy behaviour of \(H_{NB}\). 
\newline
\indent
In the limit \(\rmq\to 0\) for the GP's, one finds the following  
relations with the polarizabilities (in gaussian units) of RCS,
as discussed in section~\ref{rcsdr} \cite{Dre98c}~:
\begin{eqnarray}
&&P^{(01,01)0}(0)=-{1 \over {\alpha_{em}}}\, \sqrt{\frac{2}{3}} \,
\alpha \;, \nonumber\\
&&P^{(11,11)0}(0)=-{1 \over {\alpha_{em}}}\, \sqrt{\frac{8}{3}} \, 
\beta \;, \nonumber \\
&&P^{(01,12)1}(0)=-{1 \over {\alpha_{em}}}\, \frac{\sqrt{2}}{3} \,
\gamma_3 \;,  \nonumber\\ 
&&P^{(11,02)1}(0)=-{1 \over {\alpha_{em}}}\, \frac{2 \sqrt{2}}{3 \sqrt{3}} \,
\left( \gamma_2 + \gamma_4 \right) \;, \nonumber \\
&&P^{(01,01)1}(0)= 0 \;, \nonumber\\
&&P^{(11,11)1}(0)= 0 \;, 
\label{eq_3_47}
\end{eqnarray}
where $\alpha_{em} = 1/137.036$ is the QED fine structure constant. 

\subsection{VCS observables}

We next discuss how one can analyze $e p \to e p \gamma$ observables
to extract the 6 GP's of Eq.~(\ref{eq_3_41}). 
\newline
\indent
The VCS unpolarized squared amplitude is denoted by \({\cal M}\).  
Besides, one can consider VCS double polarization observables, which 
are denoted by \(\Delta{\cal M} (h, i) \) 
for an electron of helicity $h$, and which are defined as 
a difference of the squared amplitudes for recoil (or target) proton
spin orientation in the direction and opposite to 
the axis $i$ ($i = x, y, z$) (see Ref.~\cite{Vdh97a} for details).    
In an expansion in $\rmqp$, \({\cal M}\) and \(\Delta{\cal M}\) take 
the form
\begin{eqnarray}
\calm^{\rm exp}&=&\frac{\calm^{\rm exp}_{-2}}{\rmqp^2}
+\frac{\calm^{\rm exp}_{-1}}{\rmqp}
+\calm^{\rm exp}_0+O(\rmqp),
\nonumber\\
\dcalm^{\rm exp}&=&\frac{\dcalm^{\rm exp}_{-2}}{\rmqp^2}
+\frac{\dcalm^{\rm exp}_{-1}}{\rmqp}+\dcalm^{\rm exp}_0+O(\rmqp) \,. \;\;
\label{eq_3_50}
\end{eqnarray}
Due to the LET, the threshold coefficients 
$\calm_{-2}$, $\calm_{-1}$, $\dcalm_{-2}$, $\dcalm_{-1}$ are known. 
The information on the GP's is contained in \( \calm^{\rm exp}_0\) and
 \( \dcalm^{\rm exp}_0\). These coefficients contain a part
  which comes from the (BH+Born) amplitude 
and another one which is a linear combination of the GP's with
coefficients determined by the kinematics. 
\newline
\indent
The unpolarized observable $\calm^{\rm exp}_0$ 
was obtained by Ref. \cite{Gui95} in terms of 3 structure functions
$P_{LL}(\rmq)$, $P_{TT}(\rmq)$, $P_{LT}(\rmq)$, which 
are linear combinations of the 6 GP's, 
\begin{eqnarray}
\calm^{\rm exp}_0 -  \calm^{\rm BH+Born}_0 
&&= 2 K_2  \left\{ 
v_1 \left[ {\epsilon  P_{LL}(\rmq)  -  P_{TT}}(\rmq)\right] \right.\nonumber\\
&&\hspace{-1.5cm}\left. 
+ \left(v_2-\frac{\qt0}{\rmq}v_3\right)\sqrt {2\varepsilon \left( 
{1+\varepsilon }\right)} P_{LT}(\rmq) \right\},
\label{eq:vcsunpol}
\end{eqnarray}
where $K_2, \epsilon, \qt0, v_1, v_2, v_3$ are kinematical
quantities (for details see Ref.~\cite{Gui98}). 
\newline
\indent
The three double-polarization observables $\dcalm^{\rm exp}_0(h,i)$
($i = x,y,z$) were expressed by \cite{Vdh97a} in terms of three new
independent structure functions $P_{LT}^z(\rmq)$, $P_{LT}^{'z}(\rmq)$,
and $P_{LT}^{'\perp}(\rmq)$, 
which are also linear combinations of the 6 GP's, 
\begin{eqnarray}
&&\dcalm^{\rm exp}_0(h,z) - \dcalm^{\rm BH+Born}_0(h,z)  \nonumber\\
&& = 4 (2h) K_2 \left\{ -v_1  \sqrt {1-\varepsilon ^2}P_{TT}(\rmq)
+ v_2  \sqrt {2\varepsilon \left( {1-\varepsilon }
  \right)}P_{LT}^z(\rmq) \right. \nonumber\\
&&\left. \hspace{.7cm}
+ v_3 \sqrt {2\varepsilon \left( {1-\varepsilon } \right)}
P_{LT}^{'z}(\rmq)  \right\} ,\nonumber \\
&&\dcalm^{\rm exp}_0(h,x) - \dcalm^{\rm BH+Born}_0(h,x) \nonumber\\
&& = 4 (2h) K_2 \left\{  v_1^x 
\sqrt {2\varepsilon \left( {1-\varepsilon } \right)} \, P_{LT}^{\perp}(\rmq) 
+ v_2^x  \sqrt {1-\varepsilon ^2} \, 
P_{TT}^{\perp}(\rmq) \right. \nonumber \\
&&\hspace{.7cm} \left. + v_3^x  \sqrt {1-\varepsilon ^2} \, 
P_{TT}^{'\perp}(\rmq) + v_4^x \sqrt {2\varepsilon \left( {1-\varepsilon } 
\right)} \, P_{LT}^{' \perp}(\rmq) \right\}, \nonumber\\
&&\dcalm^{\rm exp}_0(h,y) - \dcalm^{\rm BH+Born}_0(h,y) \nonumber\\
&& = 4 (2h) K_2 \left\{ v_1^y \sqrt {2\varepsilon \left( {1-\varepsilon }
\right)} \, P_{LT}^{\perp}(\rmq) 
+ v_2^y \sqrt {1-\varepsilon ^2}\, P_{TT}^{\perp}(\rmq)
\right.\nonumber\\  
&&\hspace{.7cm} \left. + v_3^y \sqrt {1-\varepsilon ^2} \, 
P_{TT}^{' \perp}(\rmq) + v_4^y \sqrt {2\varepsilon 
\left( {1-\varepsilon } \right)} \, P_{LT}^{'\perp}(\rmq)  \right\},
\label{eq:vcspol}
\end{eqnarray}
where $v_1^x,..., v_4^x, v_1^y,..., v_4^y$ are kinematical
coefficients. The other structure functions in Eq.~(\ref{eq:vcspol}) can
be expressed in terms of 
$P_{LL}, P_{TT}, P_{LT}, P_{LT}^z, P_{LT}^{'z}, P_{LT}^{'\perp}$
\cite{Gui98}. Therefore, measuring those 6 structure functions 
amounts to determine the 6 independent GP's. 

\subsection{Results for VCS observables below pion threshold}

\begin{figure}[h]
\vspace{-0.6cm}
\epsfysize=7 cm
\centerline{\epsffile{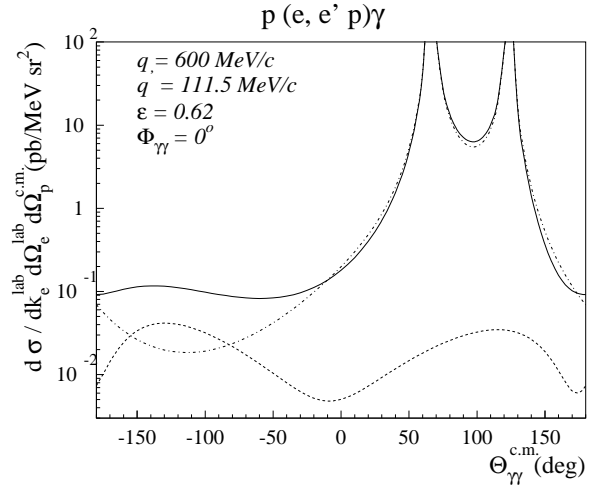}}
\vspace{-.3cm}
\caption{Results for the BH (dashed-dotted curve), 
Born (dashed curve) and BH+Born (full curve)  
p(e, e$'$ p)$\gamma$ differential cross sections in MAMI kinematics : 
q = 600 MeV/c, q$'$ = 111.5 MeV/c and $\epsilon$ = 0.62, 
as function of the c.m. angle $\Theta^{\gamma\gamma}_{\rm c.m.}$ 
between real and virtual photon, and for in-plane kinematics. 
Calculations from Ref.~\cite{Vdh96}.}
\label{fig:mamicross}
\end{figure}

In the previous sections, the observables of the 
$e p \to e p \gamma$ reaction below pion threshold were outlined, 
and it was shown how the nucleon structure effect 
can be parametrized in terms of 6 independent GP's. 
\newline
\indent
To access the GP's, the experimental strategy of 
VCS in the threshold region consists of two steps. 
First, one measures the VCS cross section at several values of 
the outgoing photon energy. At low energies, the 
measurement of the VCS observables provides a test of the LET. 
Once the LET is verified, the relative effect of the GP's  
can be extracted using Eqs.~(\ref{eq:vcsunpol},\ref{eq:vcspol}). 
\newline
\indent
The predictions for the Bethe-Heitler (BH) and Born cross sections 
below pion threshold are shown in Fig.~\ref{fig:mamicross}. 
The BH cross section has a characteristic angular shape 
and displays two ``spikes'', which occur when the direction 
of the outgoing photon coincides with either the initial or final 
electron directions. In these regions, the cross section is 
completely dominated by the BH contributions. In order to determine 
the VCS contribution, one clearly has to minimize the BH contamination
by detecting the photon in the half-plane opposite to the electron directions.

The first dedicated VCS experiment has been
performed at MAMI, and for the first time two combinations 
(see Eq.~(\ref{eq:vcsunpol})) of GP's have been determined at 
$Q^2$ = 0.33 GeV$^2$ and photon polarization $\epsilon$ = 0.62 \cite{Roc00},
\begin{eqnarray}
&&P_{LL}(Q^2) - {1 \over \epsilon} P_{TT}(Q^2) = \left(23.7 \pm 2.2
  \pm 0.6 \pm 4.3 \right)
\, {\rm GeV}^2 \, , \nonumber\\
&&P_{LT}(Q^2) = \left(-5.0 \pm 0.8 \pm 1.1 \pm 1.4 \right) \, {\rm GeV}^2 . 
\label{eq:mamiexp}
\end{eqnarray} 
VCS experiments at higher $Q^2$ (1 - 2 GeV$^2$) at JLab \cite{Bert93}, 
and at lower $Q^2$ at MIT-Bates \cite{Sha97}
have already been performed, and are under analysis at this time.
\newline
\indent
The GP's have been calculated in various 
approaches and nucleon structure models, ranging from 
constituent quark models \cite{Gui95,Pas98}, 
a relativistic effective Lagrangian model \cite{Vdh96}, 
and the linear $\sigma$-model \cite{Met97} to ChPT \cite{Hem97b,Hem00a}. 
The GP's teach us about the interplay between nucleon-core
excitations and pion-cloud effects, which are described differently in
the various models. 
We focus here on the calculation of the GP's in HBChPT 
to ${\cal O}(p^3)$, as it takes 
account of N$\pi$-loop contributions in a systematic way. 
The ${\cal O}(p^3)$ calculation yields for the two measured
combinations at $Q^2$ = 0.33 GeV$^2$ and $\epsilon = 0.62$ 
of Eq.~(\ref{eq:mamiexp}) the values \cite{Hem00a}~: 
\begin{eqnarray}
{\cal O}(p^3): \quad
P_{LL} - P_{TT}/\epsilon \;&=&\; 26.3 \;\;\;{\rm GeV}^2, \nonumber\\ 
P_{LT} \;&=&\; -5.7 \;\;\;{\rm GeV}^2, 
\end{eqnarray}
which are in astonishing agreement with the
experimentally determined values of Eq.~(\ref{eq:mamiexp}). In
particular, the ${\cal O}(p^3)$ ChPT calculation 
predicts quite large values for the spin GP's.  As for the case of the
RCS polarizabilities, the importance of the 
${\cal O}(p^4)$ corrections remains to be checked.
\newline
\indent
If one wants to extract the different polarizabilities from experiment, and in
particular in case of the spin polarizabilities, 
an unpolarized experiment is not sufficient as it gives access to  
3 independent response functions only. 
To further separate the polarizabilities, one has to resort to 
double-polarization observables. Experimentally, at the 
existing high-duty-cycle electron facilities 
with polarized electron beams such as MAMI, 
MIT-Bates and JLab, double polarization VCS experiments can be 
performed by measuring the 
recoil polarization of the outgoing nucleon with a focal plane
polarimeter. An experiment at MAMI has already been proposed \cite{d'Ho99b}.

\begin{figure}[h]
\epsfysize=7.5 cm
\centerline{\epsffile{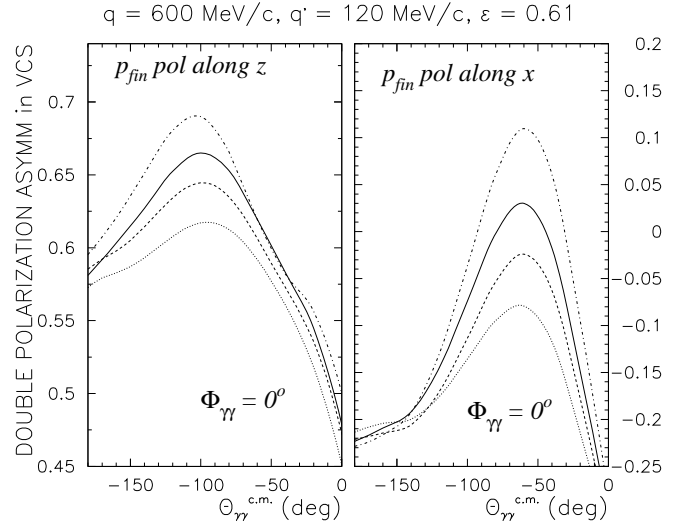}}
\vspace{-.3cm}
\caption[]{VCS double-polarization asymmetries (polarized electron, 
proton recoil polarization along either the $x$- or $z$-directions) 
in MAMI kinematics ($Q^2$ = 0.33~GeV$^2$) as function of the 
c.m. angle between real and virtual photon. 
The BH + Born result is shown by the dashed curves.  
The other curves show predictions from different model contributions
calculated in Ref.~\cite{Vdh97a}, to which we refer for details.}  
\label{fig:polmami}
\end{figure}

In Fig.~\ref{fig:polmami}, the double-polarization asymmetry of 
Eq.~(\ref{eq:vcspol}), with proton recoil polarization either 
along the $x$- or $z$- directions,  
is shown at $\rm q$ = 600 MeV/c for in-plane kinematics.
It is seen that the asymmetry yields a very large value (between 0.6 and 0.7)  
if the final proton is polarized parallel to the virtual photon.  

\subsection{Dispersion relation formalism for VCS}

At present, VCS experiments at low outgoing photon energies 
are analyzed in terms of low-energy expansions (LEXs) of Eq.~(\ref{eq_3_50}). 
In the LEX, the non-Born response of the system to the quasi-constant 
electromagnetic field of the low energetic photon is proportional to
the GP's, as expressed in Eqs.~(\ref{eq:vcsunpol},\ref{eq:vcspol}). 
As the sensitivity of the VCS cross sections to the GP's 
grows with the photon energy, it is
advantageous to go to higher photon energies, provided one can keep the
theoretical uncertainties under control when crossing the pion
threshold. The situation can be compared to RCS, 
for which one uses a dispersion relation formalism
as discussed in section~\ref{rcsdr}, 
to extract the polarizabilities at energies 
above pion threshold, with generally larger effects on the observables.  
\newline
\indent
In order to go to higher energies and to check 
the validity of LEXs at these higher energies, a dispersion
relation analysis for VCS has been developed very recently
\cite{Pas00,Vdh00b}, which will allow to extract 
the GP's from data over a larger energy range. 
The same formalism also provides for the first
time a dispersive evaluation of 4 GP's.
\newline
\indent
To set up a dispersion formalism for the VCS process,
one starts from the helicity amplitudes~:
\begin{equation}
T_{\lambda' s'; \, \lambda s} \;=\; -\,e^2 \varepsilon_\mu(q, \lambda) \, 
\varepsilon^{'*}_\nu(q', \lambda') \,
\bar u(p', s')  {\mathcal M}^{\mu \nu} u(p, s),
\label{eq:vcs_matrixele}
\end{equation}
with $e$ the electric charge, $q$ ($q'$) 
the four-vectors of the virtual (real) photon in the VCS process, 
and $p$ ($p'$) the four-momenta of the initial (final)
nucleons respectively. The nucleon helicities are
denoted by $s, s' = \pm 1/2$, and $u, \bar u$ are the nucleon spinors. 
The initial virtual photon has helicity $\lambda = 0, \pm 1$ and 
polarization vector $\varepsilon_\mu$, whereas 
the final real photon has helicity $\lambda' = \pm 1$ and 
polarization vector $\varepsilon^{'}_\nu$. The VCS process is
characterized by 12 independent helicity amplitudes 
$T_{\lambda' s'; \, \lambda s}$.   
\newline
\indent
The VCS tensor ${\mathcal M}^{\mu \nu}$ in
Eq.~(\ref{eq:vcs_matrixele}) is then expanded into a basis of 
12 independent gauge invariant tensors $\rho^{\mu \nu}_i$,  
\begin{equation}
{\mathcal M}^{\mu \nu} \;=\; \sum_{i = 1}^{12} 
\; F_i(Q^2, \nu, t) \, \rho^{\mu \nu}_i \;, 
\label{eq:vcs_nonborn}
\end{equation}
as introduced in Ref.~\cite{Dre97} 
(starting from the amplitudes of Ref.~\cite{Tar75}). 
The amplitudes $F_i$ in Eq.~(\ref{eq:vcs_nonborn}) contain all nucleon
structure information and are functions of 3 invariants for the
VCS process~: $Q^2 \equiv - q^2$, $\nu = (s - u)/(4 m_N)$  which is odd under 
$s \leftrightarrow u$ crossing, and $t$. 
The Mandelstam invariants $s$, $t$ and $u$ for VCS are defined by 
$s = (q + p)^2$, $t = (q - q')^2$, and $u = (q - p')^2$, with the
constraint $s + t + u = 2 m_N^2 - Q^2$, and $m_N$ is the nucleon mass.
\newline
\indent
Nucleon crossing combined with charge conjugation provides the
following constraints on the amplitudes $F_i$ 
\footnote{In \cite{Pas00}, 4 of the 12 amplitudes
  of \cite{Dre97} were redefined by dividing them through $\nu$, such that
  all of them are even functions of $\nu$. This simplifies the
  formalism since only one type of dispersion integrals 
  needs to be considered then.} 
at arbitrary virtuality $Q^2$ :
\begin{equation}
F_i \left( Q^2, -\nu, t \right) 
= F_i\left( Q^2, \nu, t \right) \hspace{.5cm}
(i = 1,...,12).
\label{eq:crossing}  
\end{equation}
With the choice of the tensor basis of Ref.~\cite{Dre97}, 
the resulting non-Born amplitudes $F^{NB}_i$ ($i$ = 1,...,12)
are free of all kinematical singularities and constraints. 
\newline
\indent
Assuming further analyticity and
an appropriate high-energy behavior, the non-Born amplitudes 
$F^{NB}_i(Q^2, \nu, t)$ fulfill 
unsubtracted dispersion relations (DR's) 
with respect to the variable $\nu$ at
fixed $t$ and fixed virtuality $Q^2$~:
\begin{equation}
{\mathrm Re} F_i^{NB}(Q^2, \nu, t) = 
{2 \over \pi}  {\mathcal P} \int_{\nu_{thr}}^{+ \infty} d\nu'  
{{\nu' \, {\mathrm Im}_s F_i(Q^2, \nu',t)} \over {\nu'^2 - \nu^2}},
\label{eq:vcs_unsub} 
\end{equation}
with ${\mathrm Im}_s F_i$ the discontinuities 
across the $s$-channel cuts of the VCS process. 
Since pion production is the first inelastic channel, 
$\nu_{thr} = m_\pi + (m_\pi^2 + t/2 + Q^2/2)/(2 m_N)$, 
where $m_\pi$ denotes the pion mass. 
\newline
\indent
The unsubtracted DR's of Eq.~(\ref{eq:vcs_unsub}) require 
that at sufficiently high energies ($\nu \rightarrow \infty$ 
at fixed $t$ and fixed $Q^2$) the
amplitudes ${\mathrm Im}_s F_i(Q^2,\nu,t)$ ($i$ = 1,...,12) 
drop fast enough such that the
integrals are convergent and the contributions from
the semi-circle at infinity can be neglected. 
It turns out that for two amplitudes, $F_1$ and $F_5$, 
an unsubtracted dispersion integral as in Eq.~(\ref{eq:vcs_unsub})
does not exist \cite{Pas00}, whereas the other 10 amplitudes can 
be evaluated through unsubtracted dispersion integrals. 
This situation is similar as for RCS, 
where 2 of the 6 invariant amplitudes cannot be evaluated 
by unsubtracted dispersion relations either \cite{Lvo97}.
\newline
\indent 
The unsubtracted DR formalism for VCS also allows to predict 4 of the
6 GP's. The appropriate limit in the definition of the GP's is 
$\rmqp \to 0$ at finite $\rmq$ (see Eq.~(\ref{eq_3_35})), 
which corresponds in terms of VCS invariants to 
$\nu \to 0$ and $t \to -Q^2$ at finite $Q^2$. 
One can therefore express the GP's in terms of
the VCS amplitudes $F_i$ at the point $\nu = 0$, $t = -Q^2$ at
finite $Q^2$, denoted in the following as~:
$\bar F_i(Q^2) \;\equiv\; F_i^{NB} \left(Q^2, \nu = 0, t = - Q^2 \right)$.
The relations between the GP's and the $\bar F_i(Q^2)$ can be found in
Ref.~\cite{Dre97}. 
From the high-energy behavior for the VCS invariant amplitudes, 
it follows that one can evaluate the $\bar F_i$ (for $i \neq$ 1, 5) 
through the unsubtracted DR's
\begin{equation}
\bar F_i(Q^2) \;=\; 
{2 \over \pi} \; \int_{\nu_{thr}}^{+ \infty} d\nu' \; 
{{{\mathrm Im}_s F_i(Q^2, \nu',t = - Q^2)} \over {\nu'}}\;.
\label{eq:vcs_sumrule} 
\end{equation} 
Unsubtracted DR's for the GP's will 
therefore hold for those combinations of GP's that do not  
depend upon the amplitudes $\bar F_1$ and $\bar F_5$ 
\footnote{$\bar F_5$ can appear however in the combination 
$\bar F_5 + 4 \, \bar F_{11}$, in which the $\pi^0$-pole 
drops out, and which has a high-energy behavior 
leading to a convergent integral (see Ref.~\cite{Pas00}).}. 
Among the 6 GP's, the following 4 combinations of GP's were found in
Ref.~\cite{Pas00}~:
\begin{eqnarray}
&&P^{\left(0 1, 0 1\right)0} + {1 \over 2}  
P^{\left(1 1, 1 1\right)0} = 
{{-2} \over {\sqrt{3}}}\,\left( {{E + m_N} \over
    E}\right)^{1/2} m_N\,\tilde q_0\, \nonumber\\
&&\hspace{.8cm} \times 
\left\{ {{\rmq^2} \over {\tilde q_0^2}}\, \bar F_2 + 
\left( 2 \, \bar F_6 + \bar F_9 \right) - \bar F_{12} \right\}, 
\label{eq:gpdisp1} \\
&&P^{\left(0 1, 0 1\right)1} = 
{1 \over {3 \sqrt{2}}}\,\left( {{E + m_N} \over E}\right)^{1/2} 
\,\tilde q_0\, \nonumber\\
&&\hspace{.8cm}\times 
\left\{ \left( \bar F_5 + \bar F_7 + 4\, \bar F_{11} \right) 
+ 4 \, m_N \, \bar F_{12} \right\}, 
\label{eq:gpdisp2} \\
&&P^{\left(0 1, 1 2\right)1} - {1 \over {\sqrt{2} \, \tilde q_0}}  
P^{\left(1 1, 1 1\right)1} = 
{1 \over {3}} \left( {{E + m_N} \over E}\right)^{1/2} 
{{m_N \, \tilde q_0} \over {\rmq^2}} \nonumber\\
&&\hspace{.8cm}\times
\left\{ \left( \bar F_5 + \bar F_7 + 4\, \bar F_{11} \right) 
+ 4 \, m_N \left( 2 \, \bar F_6 + \bar F_9 \right) \right\}, 
\label{eq:gpdisp3} \\
&&P^{\left(0 1, 1 2\right)1} +  
{{\sqrt{3}} \over {2}}  P^{\left(1 1, 0 2\right)1} =
{1 \over {6}} \left( {{E + m_N} \over E}\right)^{1/2} \, 
{{\tilde q_0} \over {\rmq^2}} \nonumber \\
&&\hspace{.8cm} \times 
\left\{ \tilde q_0 \left( \bar F_5 + \bar F_7 + 4\, \bar F_{11} \right) 
+ 8 \, m_N^2 \left( 2 \, \bar F_6 + \bar F_9 \right) \right\}, \;\;\;\;\;
\label{eq:gpdisp4} 
\end{eqnarray}
where $E = \sqrt{\rmq^2 + m_N^2}$ denotes 
the initial proton c.m. energy, and  
$\tilde q_0 = m_N - E$ the virtual photon c.m. energy in the limit 
$\rmqp$ = 0. Unfortunately, the 4 combinations of GP's of 
Eqs.~(\ref{eq:gpdisp1})-(\ref{eq:gpdisp4}) can at present not yet
be compared with the data. In particular, the only unpolarized experiment 
\cite{Roc00} measured two structure functions which 
cannot be evaluated in an unsubtracted DR formalism, 
as they contain in addition to 
$P^{\left(0 1, 0 1\right)0} + 1/2 P^{\left(1 1, 1 1\right)0}$ 
of Eq.~(\ref{eq:gpdisp1}), which
is proportional to $\alpha + \beta$ at $Q^2$ = 0, also the
generalization of $\alpha - \beta$. 
\newline
\indent
The 4 combinations of GP's on the {\it lhs} of 
Eqs.~(\ref{eq:gpdisp1})-(\ref{eq:gpdisp4}) can then be evaluated by 
unsubtracted DR's, from the dispersion integrals
of Eq.~(\ref{eq:vcs_sumrule}) for the $\bar F_i(Q^2)$. 
To this end, the imaginary parts ${\mathrm Im}_s F_i$ 
in Eq.~(\ref{eq:vcs_sumrule}) have to be calculated by use of unitarity. 
For the VCS helicity amplitudes of Eq.~(\ref{eq:vcs_matrixele})
(denoted for short by $T_{fi}$), the unitarity equation reads~:
\begin{equation}
\label{s-unit}
2\,{\rm Im}_s\,T_{fi}=
\sum_X (2\pi)^4 \delta^4(P_X-P_i)T^{\dagger}_{X f }\,T_{X i} \;,
\end{equation}
where the sum runs over all possible intermediate states 
$X$ that can be formed. 
In Ref.~\cite{Pas00}, the dispersion integrals of
Eq.~(\ref{eq:vcs_sumrule}) were saturated 
by the dominant contribution of the $\pi N$
intermediate states. For the pion photo- and electroproduction
helicity amplitudes in the range $Q^2 \leq$ 0.5 GeV$^2$, the   
phenomenological analysis of MAID \cite{maid00} was used, 
which contains both resonant and non-resonant pion production mechanisms.
\begin{figure}[h]
\epsfxsize=8.25 cm
\epsfysize=8.25 cm
\vspace{-.25cm}
\centerline{\epsffile{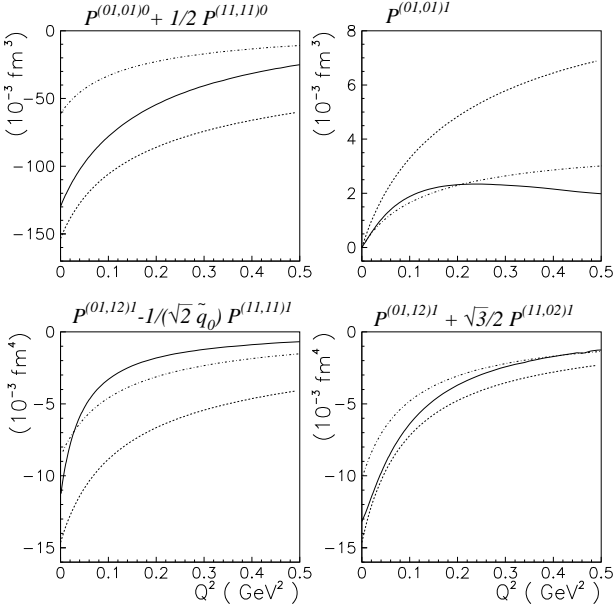}}
\vspace{-.2cm}
\caption[]{\small Dispersion results \cite{Pas00} for 4 of the generalized 
polarizabilities of the proton 
(full curves), compared with results of $O(p^3)$ 
HBChPT \cite{Hem97b,Hem00a} (dashed curves) and the linear $\sigma$-model
\cite{Met97} (dashed-dotted curves).}
\label{fig:polcomb}
\end{figure}

In Fig.~\ref{fig:polcomb}, the results for the 4 combinations of GP's 
of Eqs.~(\ref{eq:gpdisp1})-(\ref{eq:gpdisp4}) are shown
in the DR formalism, and compared to the results of the 
$O(p^3)$ heavy-baryon chiral perturbation theory (HBChPT) \cite{Hem97b,Hem00a}
and the linear $\sigma$-model \cite{Met97}. 
The $\pi N$ contribution to the sum 
$P^{\left(0 1, 0 1\right)0}$ + $1/2 P^{\left(1 1, 1 1\right)0}$ 
gives only about 80\% of
the Baldin sum rule \cite{Bab98}, because of a non-negligible
high-energy contribution (of heavier intermediate states) to the  
photoabsorption cross section entering the sum rule, 
which is not estimated here. On the other hand, for the 3 combinations
of spin polarizabilities of Eqs.~(\ref{eq:gpdisp2})-(\ref{eq:gpdisp4}), 
the dispersive estimates with $\pi N$ states are expected to
provide a rather reliable guidance. By comparing the DR results with those 
of HBChPT at $O(p^3)$, one remarks a rather good agreement for  
$P^{\left(0 1, 1 2\right)1}$ + $\sqrt{3}/2 P^{\left(1 1, 0 2\right)1}$,  
whereas for the GP's $P^{\left(0 1, 0 1\right)1}$ and 
$P^{\left(0 1, 1 2\right)1}$ - $1/(\sqrt{2} \, \tilde q_0) 
P^{\left(1 1, 1 1\right)1}$,
the dispersive results drop much faster with $Q^2$. This trend
is also seen in the relativistic linear $\sigma$-model, which 
takes account of some higher orders in the chiral expansion. 
It remains to be checked how the ${\cal O}(p^4)$ corrections 
in HBChPT change this comparison with the DR estimates. 
\newline
\indent
To complete the DR formalism for VCS, one further needs to 
construct the VCS amplitudes $F_1$ and $F_5$, for which the unsubtracted 
dispersion integrals of Eq.~(\ref{eq:vcs_unsub}) do not converge. 
One strategy is to proceed in an analogous way as 
has been proposed in Ref.~\cite{Lvo97} in the case of RCS.  
The unsubtracted dispersion integrals for $F_1$ and $F_5$ 
are evaluated along the real $\nu$-axis in a finite range 
$-\nu_{max}\leq\nu\leq+\nu_{max}$ (with $\nu_{max}\approx$ 1.5~GeV).   
The integral along a semi-circle of finite radius $\nu_{max}$ 
in the complex $\nu$-plane is described by  
the asymptotic contribution $F_i^{as}$, 
which is parametrized by $t$-channel poles (e.g. for $Q^2$ = 0,
$F_1^{as}$ corresponds to $\sigma$-exchange, and $F_5^{as}$ 
to $\pi^0$-exchange). 
\newline
\indent
A full study of VCS observables within such a dispersion
formalism, including a parametrization of the two asymptotic
contributions, is presently underway \cite{Dre00c}.
This will yield a formalism to extract the nucleon GP's over a 
larger range of energies from both unpolarized and polarized VCS data.

\section{Compton scattering at large momentum transfer and the nucleon
distribution amplitude}
\label{larget}

\subsection{Introduction}

\begin{figure}[h]
\epsfxsize=8.25cm
\vspace{-.2cm}
\centerline{\epsffile{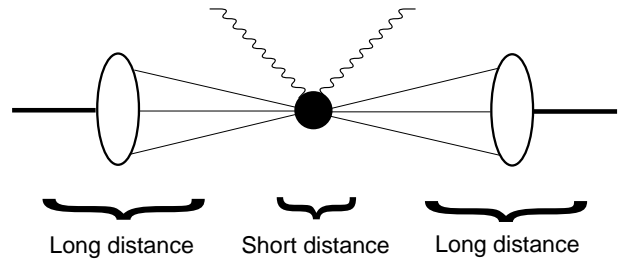}} 
\vspace{0cm}
\caption{\small Factorization of the RCS scattering amplitude in the hard 
scattering regime.}
\label{fig:facto}
\end{figure}

Besides the low energy region, as discussed in section~\ref{rcsdr}, RCS
will also provide access to information on the partonic structure of the
nucleon at sufficiently large momentum transfer. 
\newline
\indent
This regime is defined  by requiring that all three 
Mandelstam variables ($s,\,t,\ u$) be
large with respect to a typical hadronic scale, say 1 GeV. In this case there
is a prejudice (actually proven in the case of elastic electron scattering
\cite{Ste97}) that the amplitude factorizes in a soft non-perturbative part, 
the distribution amplitude, and a
hard  scattering kernel  which is calculable from perturbative QCD (PQCD). 
Because of asymptotic freedom, 
the perturbative approach must be to some degree relevant to
the hard scattering regime. However, since the binding of the quarks
and gluons in
the hadrons is a long distance, non-perturbative effect, the description of the
reaction requires a consistent analysis of both large and small scales.
When the reaction is hard enough, 
the relative velocities of the participating particles are nearly
lightlike. Time dilatation increases the lifetime of the quantum configurations
which build the hadron. As a result, 
the partonic content, as seen by the other 
particles, is frozen. Moreover, due to the apparent contraction 
of the hadron size,
the time during which momentum can be exchanged is decreased. Therefore one
expects a lack of coherence between the long-distance confining effects and the
short distance reaction. This incoherence between the soft and hard physics is
the origin of the factorization which is illustrated in Fig.~\ref{fig:facto}.

\subsection{Factorization and the nucleon distribution amplitude}

\begin{figure}[h]
\epsfxsize=6.cm
\vspace{.25cm}
\centerline{\epsffile{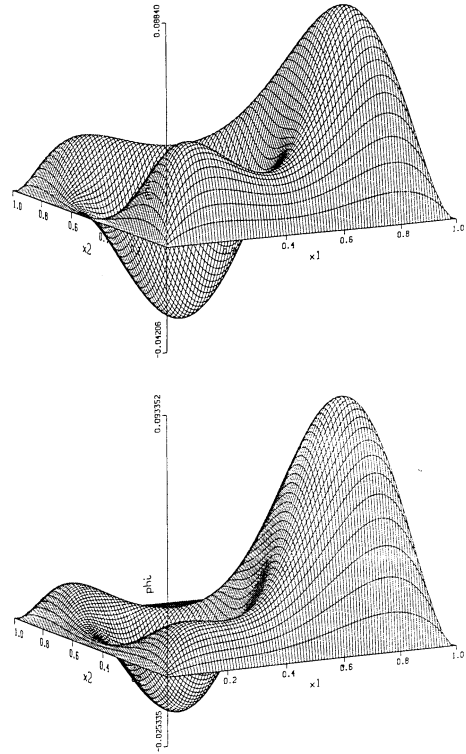} }
\caption{\small Model distribution amplitudes for the nucleon : 
KS (upper figure) and COZ (lower figure), as function of the 
valence quark momentum fractions $x_1$ and $x_2$ ($x_1 + x_2 + x_3$ = 1).}
\label{fig:PQCD}
\end{figure}

The calculation of the RCS amplitude at large momentum transfer, 
follows the Brodsky-Lepage 
formalism \cite{Bro80}, which leads to the factorized expression~:
 \begin{eqnarray}
\label{eq:fact}
&&T(\lambda ',h_N',\lambda,h_N) \;=\; \nonumber\\
&&\int dx_i \, dy_j \, \phi_N ^* (y_j) \,  
T_{H}(\lambda ',h_N',y_j,\lambda,h_N,x_i; s, t) \, \phi_N (x_i), \nonumber\\
&&
\end{eqnarray}
where ($x_i,\,y_i$) are the momentum fractions of the quarks 
in the initial and final nucleon respectively, 
$T_H$ is the hard scattering kernel and 
$\phi_N$ is the distribution amplitude (DA). 
The evaluation of Eq.~(\ref{eq:fact}) requires a four-fold 
convolution integral since there are 
two constraint equations $(x_1 + x_2 + x_3 = 1$ and $y_1 + y_2 + y_3 = 1)$.
In  Eq.~(\ref{eq:fact})  a sufficiently large 
momentum transfer is assumed in order to neglect the transverse 
momentum dependence of the partons in the 
hard scattering amplitude $T_H$. In this limit, the integration over the 
transverse momenta $\vec{k}_{\perp i}$  
(where $\sum _i \vec{k}_{\perp i} = \mathbf 0$) 
acts only on the valence wavefunction  
\begin{equation}
\Psi_V ( x_1, x_2, x_3; \vec{k}_{\perp 1}, \vec{k}_{\perp 2},
\vec{k}_{\perp 3} )\;, 
\end{equation}
which is the amplitude of the 
three quark state in the Fock expansion of the proton:
\begin{eqnarray}
\mid P > &=& \Psi_V  \mid q q q >
 + \Psi_{q \overline q }  \mid q q q, q {\overline q } >
 + \Psi_g  \mid q q q, g > \nonumber\\
&& +...
\end{eqnarray}
This valence wave function $\Psi_V$ integrated up to a scale $\mu$ 
(which separates the soft and hard parts of the wavefunction) defines the 
DA which appears in Eq.~(\ref{eq:fact}) : 
\begin{equation}
\label{eq:wavefunction}
  \phi_N( x_i,\mu ) = \int ^\mu d^2 \vec{k}_{\perp i} \;
                                 \Psi_V ( x_i; \vec{k}_{\perp i} )\;.
\end{equation}
For $ \mu $ much larger than the average value of the transverse momentum 
in the proton, this function $ \phi_N $ depends only weakly on $ \mu $ 
\cite{Bro80} and this dependence can be neglected.
\newline
\indent
The interest of the formalism is that the distribution amplitude is universal,
that is  independent of the particular reaction considered. 
Several  distribution amplitudes have 
been modeled  using QCD sum rules \cite{Che84,Che89,Kin87}.
They have a characteristic shape and 
predict that in a proton, the $u$-quark with helicity along the 
proton helicity carries about 2/3 of its longitudinal momentum (see
Fig.~\ref{fig:PQCD} ).
\newline
\indent
For the computation of the hard scattering amplitude $T_H$ (black
circle in Fig.~\ref{fig:facto}), 
the leading order PQCD contribution corresponds 
to the exchange of the minimum number of 
gluons (two in the present case) between the three quarks. 
The number of diagrams grows rapidly with the 
number of elementary particles involved in the reaction 
(42 diagrams for the nucleon form factor, 
336 diagrams in the case of real or virtual Compton scattering). 
Despite the large number of diagrams, the calculation of $T_H$ 
is a parameter free calculation once the scale 
$\Lambda_{QCD} \approx$ 200 MeV 
in the strong coupling $\alpha_s (Q^2)$ is given. 
Note that configurations with more than three
valence quarks are a priori allowed but since this implies the exchange
of more hard gluons, 
the corresponding contribution is suppressed by powers of $1/t$.

There are two characteristic features of the Brodsky-Lepage model which are
almost direct consequences of QCD: 
the dimensional counting rules \cite{Bro73} 
and the conservation of hadronic helicities \cite{Bro81}. 
The latter feature implies that any helicity flip amplitude is zero and, hence,
any single spin asymmetry too. The helicity sum rule is a consequence of 
utilizing the collinear approximation and of dealing with (almost) massless 
quarks which conserve their helicities when interacting with gluons. Whereas 
the dimensional counting rules are in reasonable agreement with experiment, 
the helicity sum rule seems to be violated even at moderately large 
momentum transfers. The prevailing opinion is that these phenomena cannot 
be explained in terms of perturbative QCD (see, for example, 
Ref.~\cite{Siv89}), but rather are generated by an interplay of 
perturbative and non-perturbative physics.
\newline
\indent
An interesting aspect of real and virtual Compton 
scattering is that these are the simplest processes in which 
the integrals over the longitudinal momentum fractions yield imaginary parts.
The reason is that, as in any scattering process, 
there are kinematical regions where internal quarks and 
gluons can go on their mass shell. 
The appearance of imaginary parts to leading order in $\alpha_s$ is a 
non-trivial prediction of PQCD, which should be tested experimentally. 
As discussed in \cite{Gui98}, 
the ($e, e' \gamma $) reaction  with polarized incoming
electrons seems to be a good candidate for this investigation.
\newline
\indent
In contrast to the PQCD (or hard scattering) approach to RCS, 
it was argued in Refs.~\cite{Rad98,Die99} that wide angle
Compton scattering at accessible energies is described by a competing
mechanism, in which the large momentum transfer is absorbed on a single
quark and shared by the overlap of high-momentum components in the
soft wave function. This so-called soft-overlap mechanism 
gives a purely real amplitude, therefore displaying a different signature 
than the PQCD amplitude.  
The transition from such a soft-overlap mechanism to the 
perturbative, hard scattering approach when increasing the momentum transfer  
is an open question for a reaction such as wide angle Compton
scattering. It is hoped that future experiments can shed light on this
transition.

\subsection{Results for RCS in PQCD}

\begin{figure}[h]
\epsfxsize=8.25cm
\vspace{-.3cm}
\centerline{\hspace{.2cm} \epsffile{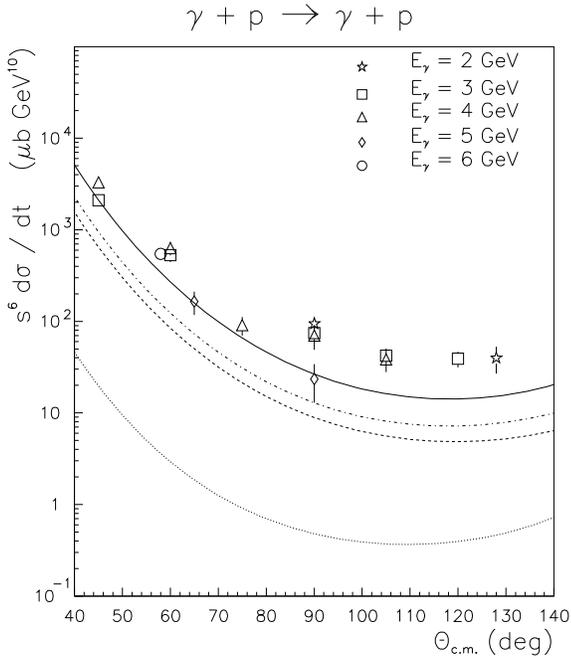} }
\vspace{-.35cm}
\caption{\small Unpolarized Compton cross section on the proton for different
nucleon DA's : KS (full line), COZ (dashed-dotted line), CZ (dashed line) 
and asymptotic DA (dotted line). 
Calculations are from Ref.~\cite{Vdh97b}, 
where also the references of the data can be found.}
\label{fig:unpol}
\end{figure}

\begin{figure}[h]
\epsfxsize=9.cm
\vspace{-.7cm}
\centerline{\hspace{.25cm}\epsffile{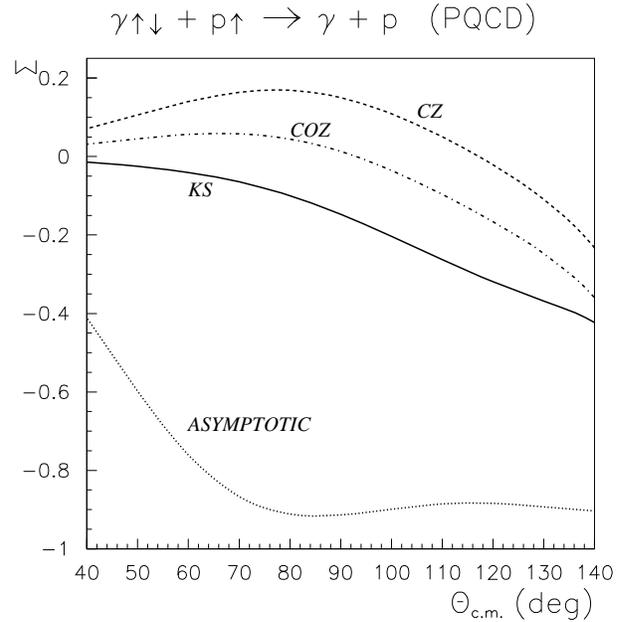} }
\vspace{-.45cm}
\caption{\small PQCD calculations of the photon asymmetry on a
  polarized proton target for Compton scattering. 
Results (from Ref.~\cite{Vdh97b}) 
are shown for different DA's as indicated on the curves.}
\label{fig:asymm}
\end{figure}

The leading order PQCD prediction for RCS at large momentum transfer
has been calculated several times in the literature 
\cite{Far90a,Far90b,Kro91,Vdh97b,Broo00}. 
\newline
\indent
The first step in such a calculation consists of
evaluating the 336 diagrams entering the 
hard scattering amplitude $T_H$ for RCS. Next, the four-fold 
convolution integral of Eq.~(\ref{eq:fact}) has to be performed to obtain 
the Compton helicity amplitudes. The numerical integration 
requires some care because the quark and/or gluon propagators can go 
on-shell which leads to (integrable) singularities. 
The different numerical implementations of these singularities are  
probably the reason of the different results obtained in the
literature. 
\newline
\indent
In Refs.~\cite{Far90a,Far90b}, the propagator singularities were integrated by 
taking a finite value for the imaginary part +i$\epsilon$ 
of the propagator. The behavior of the result was then 
studied by decreasing the value of $\epsilon$. 
To obtain convergence with a practical number of samples 
in the Monte Carlo integration performed in \cite{Far90a,Far90b}, 
the smallest feasible  value for $\epsilon$ was $\epsilon \approx 0.005$. 
In Ref.~\cite{Kro91}, the propagator singularities were integrated by 
decomposing the propagators into a principal value (off-shell) part and 
an on-shell part. 
Both methods were implemented and compared in Ref.~\cite{Vdh97b}, 
and it was found that the +i$\epsilon$ method yields 
differences of the order of 10\% for every diagram
as compared with the result of the principal value method. 
It is not surprising that, 
when summing hundreds of diagrams, an error of 10\% on every diagram 
can easily be amplified due to the interference between the diagrams.
\newline
\indent 
To have confidence in the evaluation of the convolution of 
Eq.~(\ref{eq:fact}), the principal value integration method 
was compared in Ref.~\cite{Vdh97b} with a third independent method. 
This third method starts from the observation that the diagrams can 
be classified into four categories depending upon the number of 
propagators which can develop singularities : in the present case this number 
is 0, 1, 2 or 3. Besides the trivial case of zero singularities which can
be integrated immediately, the diagrams with one or two propagator 
singularities can be integrated by performing a contour integration in the 
complex plane for one of the four integrations. For the most difficult case 
of three propagator singularities, it was found to be possible 
to evaluate the diagram by 
performing two contour integrations in the complex plane. In doing so, one 
achieves quite a fast convergence because the integrations along the real axis
are replaced by integrations along semi-circles in the complex plane which 
are far from the propagator poles.
This method was compared with the principal value integration 
method, and the same results were found up to 0.1\% 
for each type of singularity \cite{Vdh97b}. 
The principal value method was however found to converge much slower
and is more complicated to implement, 
especially for the case with three singularities 
due to the coupled nature of the three principal value integrals.
\newline
\indent
Comparing the results of Ref.~\cite{Vdh97b} with those of Ref.~\cite{Kro91}, 
a rather good agreement was found for 
all helicity amplitudes, except for the helicity amplitude where both
photon and proton helicities are positive, 
in which case both calculations differ strongly. 
Very recently, the PQCD calculation for RCS at large momentum transfer has
been recalculated again in Ref.~\cite{Broo00}, 
by also performing convolution integrals through contour integrations 
in the complex plane.  
The authors of Ref.~\cite{Broo00} also find a strong 
difference with the results of \cite{Kro91} for the same helicity
amplitude, where both photon and proton helicities are positive. 
Furthermore, in the angular region around $90^o$, where the PQCD formalism 
is supposed to be applicable, the authors 
of Ref.~\cite{Broo00} find a good agreement with the calculations of
Ref.~\cite{Vdh97b}, keeping in mind that there is an overall
normalization uncertainty in these PQCD calculations for RCS,
associated with $\alpha_s$ and the valence quark wave function normalization. 
The remaining difference between the results of Refs.~\cite{Vdh97b}
and those of Ref.~\cite{Broo00} seems to be isolated to a single helicity
amplitudes and appears for backward scattering angles. 
We therefore limit ourselves in the following discussion 
to the results in the angular region around $90^o$ where the calculations of 
Refs.~\cite{Vdh97b} and \cite{Broo00} are in good agreement,
and which is the most relevant region for the PQCD calculation 
as it corresponds to the largest momentum transfer for a given photon energy. 
\newline
\indent
In Figs.~\ref{fig:unpol} and \ref{fig:asymm}, the PQCD calculations for
RCS are shown for several model DA's denoted 
as CZ \cite{Che84}, COZ \cite{Che89}, KS \cite{Kin87}, and 
the asymptotic DA.
\newline
\indent
The highest energy data which exist for real Compton scattering 
were taken around 5 GeV and are shown in Fig.~\ref{fig:unpol}.
Although the energy at which these experiments were performed 
is probably too low to justify a PQCD calculation, 
the comparison with these data is nevertheless shown in Fig.~\ref{fig:unpol} 
for illustrative purposes. 
The normalization of the calculations shown at these very low
scales corresponds to using a frozen coupling constant,  
with $\alpha_s \approx 0.5$. 

One first notices that the hard scattering amplitude for RCS 
has the $s$-dependence ($ T \sim s^{-2}$) 
which leads to the QCD scaling laws \cite{Bro73}, that is 
${{d \sigma} \over {d t}} \sim s^{-6}$ for Compton scattering or VCS. 
The unpolarized real Compton differential cross section 
(multiplied by the scaling factor $s^6$) is shown 
in Fig.~\ref{fig:unpol} as function of the photon c.m. angle. 
It is observed that the result 
with the asymptotic DA ($\sim 120 \, x_1 x_2 x_3$) 
is more than one decade below the 
results obtained with the amplitudes KS, COZ, and CZ, 
motivated by QCD sum rules. 
The results with KS, COZ and CZ show a similar characteristic angular 
dependence which is asymmetric around 90$^o$. Note that in the forward and 
backward directions, which are dominated by diffractive mechanisms, 
a PQCD calculation is not reliable. 
Comparing the results obtained with KS, COZ and CZ, one notices that 
although these DA's have nearly the same lowest moments, 
they lead to differences of a factor of two in the Compton 
scattering cross section. Consequently, this observable is sensitive 
enough to distinguish between various distribution amplitudes, 
provided, of course, one is in the regime where the hard scattering
mechanism dominates. 
\newline
\indent
In Fig.~\ref{fig:asymm}, the polarized Compton cross sections 
are shown for the two helicity states of the photon and for a target 
proton with positive helicity. One remarks that for all DA's  
there is a marked difference both in magnitude and 
angular dependence between the cross sections for the two photon helicities. 
Consequently, the resulting photon asymmetry $ \Sigma $, defined as
\begin{equation}
 \Sigma_\uparrow = \frac{\frac{d\sigma}{dt}(\uparrow,\lambda=1)
                 -\frac{d\sigma}{dt} (\uparrow,\lambda=-1)}
                 {\frac{d\sigma}{dt}(\uparrow,\lambda=1)
                 +\frac{d\sigma}{dt} (\uparrow,\lambda=-1)},
\label{eq:rcsasymm}
\end{equation}
where $ \lambda $ is the helicity of the incoming photon and 
 $ \uparrow $ denotes a positive hadron helicity,  
changes sign for the DA's KS, COZ, and CZ 
for different values of $\Theta_{\rm c.m.}$ 
as shown in Fig.~\ref{fig:asymm}. 
It is seen that the asymptotic DA on the other hand yields a very
 large, negative asymmetry around $90^o$. 
Therefore, it was suggested in Ref.~\cite{Vdh97b} that the 
photon asymmetry might be a particularly 
useful observable to distinguish between nucleon distribution amplitudes.   
The predicted sensitivity of the asymmetry to the nucleon DA can be used in 
the extraction of a DA from Compton scattering data in the scaling region. 
In Ref.~\cite{Vdh97b}, a procedure was outlined to extract a DA from 
Compton data in a model independent way by first expanding the DA 
in a set of basis functions and then using the angular information 
of the cross sections to fit the expansion coefficients. It was seen that 
the precision for these coefficients is greatly improved when 
one measures both unpolarized cross sections and photon asymmetries. 

A first dedicated experiment to measure the RCS differential cross section and 
the asymmetry of Eq.~(\ref{eq:rcsasymm}) for 
$\Theta_{\rm c.m.}$ around $90^o$, 
and for a real photon energy of 6 GeV, is planned at JLab \cite{Woj99}. 
In particular, it will be interesting
to see if one approaches the PQCD result at these ``lower'' energies, 
and to study the interplay with soft-overlap type contributions
for RCS as proposed in Refs.~\cite{Rad98,Die99}. 
\newline
\indent
RCS experiments using a real photon energy in the 15 GeV range, 
might be feasible e.g. at the HERA ring in the foreseeable future 
\cite{d'Ho96,D'An97} and might open up prospects to study the 
nucleon valence wave function in a direct way. 

\newpage 

\section{Deeply virtual Compton scattering and skewed parton distributions}
\label{dvcs}

\subsection{Introduction}

Much of the internal structure of the nucleon has been revealed during the last
two decades through the {\it inclusive} scattering of high energy leptons
on the nucleon in the Bjorken -or ``Deep Inelastic Scattering'' (DIS)-
regime (where the photon virtuality $Q^2$ is very large, and  
$x_{B}=Q^{2}/2 p.q$ finite). 
{\it Unpolarized} DIS experiments have mapped out the quark and gluon
distributions in the nucleon, while {\it polarized} DIS experiments
have shown that only a small fraction of the nucleon spin is carried
by the quarks. 
This has stimulated new investigations to understand the  nucleon spin.
\newline
\indent 
With the advent of the new generation of high-energy, high-luminosity
lepton accelerators combined with large acceptance spectrometers, a
wide variety of {\it exclusive} processes in the Bjorken regime 
are considered as experimentally accessible. 
In recent years, a unified theoretical description of such processes
has emerged through a formalism introducing a new type of parton 
distributions, commonly denoted as skewed parton distributions 
(SPD's) \cite{Ji97,Rad96a}. These SPD's 
are generalizations of the parton distributions measured in DIS. 
It has been shown that these SPD's, which parametrize the 
structure of the nucleon, allow one to describe, in leading
order perturbative QCD (PQCD), various exclusive processes in the near forward
direction, where the momentum transfer to the nucleon is small. 
Such non-forward processes were already considered in the literature a
longer time ago, see e.g. \cite{Wat82,Bar82,Dit88,Jai93,Mul94}. 
The most promising of these non-forward 
hard exclusive processes are deeply virtual Compton scattering (DVCS)
and longitudinal electroproduction of vector or pseudoscalar mesons 
at large $Q^2$.

\subsection{Definitions and modelizations of skewed parton distributions}

\begin{figure}[ht]
\hspace{-.5cm}
\epsfxsize=8 cm
\epsfysize=6.5 cm
\centerline{\epsffile{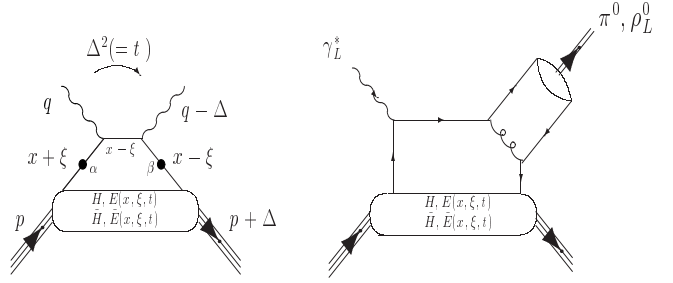}}
\vspace{-2.5cm}
\caption[]{Leading order diagrams for DVCS (left) and for longitudinal
  electroproduction of mesons (right).}
\label{fig:handbags}
\end{figure}

\begin{figure}[ht]
\epsfysize=10.5cm
\centerline{\hspace{.3cm} \epsffile{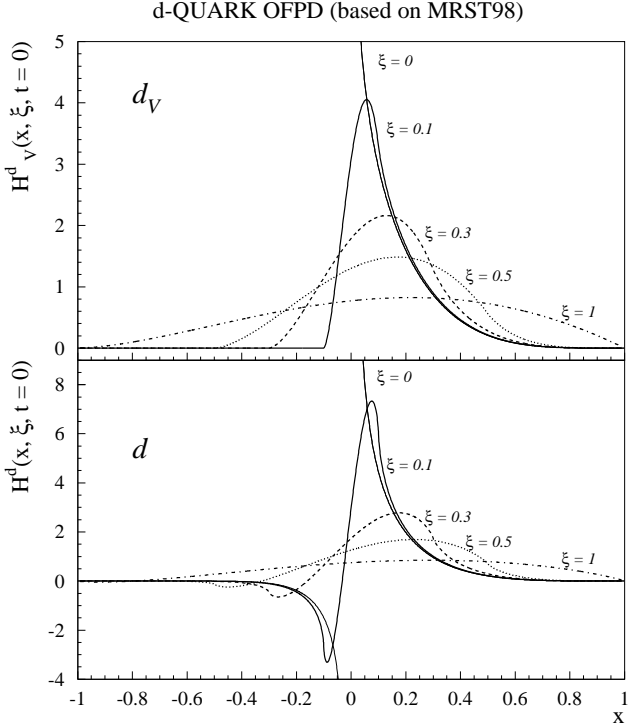}}
\vspace{-0.8cm}
\caption[]{\small $\xi$ dependence of the SPD $H^d$ at $t = 0$ 
with the ansatz (based on the MRST98 \cite{MRST98} quark
distributions) used in Ref.~\cite{Vdh99}.  
Upper panel~: valence $d$-quark SPD, lower panel~: total $d$-quark
SPD. The thin lines ($\xi = 0$) correspond with the ordinary
$d$-quark distributions.}
\label{fig:ofpdxi}
\end{figure}

The leading order PQCD diagrams for DVCS
and hard meson electroproduction are of the type as shown in 
Fig.~\ref{fig:handbags}. The hard scale in Fig.~\ref{fig:handbags} is
the photon virtuality $Q^2$, which should be large (of the order of several
GeV$^2$), so as to be in the Bjorken regime. 
It has been proven \cite{Ji97,Rad96a} 
that the leading order DVCS amplitude in
the forward direction can be factorized in a hard scattering part 
(which is exactly calculable in PQCD) and a soft, nonperturbative nucleon 
structure part as illustrated on the left panel of 
Fig.~\ref{fig:handbags}. 
The nucleon structure information  
can be parametrized, at leading order, in 
terms of four (quark helicity conserving) generalized structure
functions. These functions are the SPD's 
denoted by $H, \tilde H, E, \tilde E$
which depend upon three variables : $x$, $\xi$ and $t$.
The light-cone momentum \footnote{using the definition 
$a^{\pm} \equiv 1/\sqrt{2} (a^0 \pm a^3)$ for the light-cone
components} fraction $x$ is defined by $k^+ = x P^+$,
where $k$ is the quark loop momentum and  
$P$ is the average nucleon momentum ($P = (p + p')/2$, where $p (p')$
are the initial (final) nucleon four-momenta respectively). 
The skewedness variable $\xi$ is 
defined by $\Delta^+ = - 2 \xi \, P^+$, where $\Delta = p' - p$ is the 
overall momentum transfer in the process, and where 
$2 \xi \rightarrow x_B/(1 - x_B/2)$ in the Bjorken limit. 
Furthermore, the third variable entering the SPD's 
is given by the Mandelstam invariant $t = \Delta^2$, being the total
squared momentum transfer to the nucleon. 
In a frame where the virtual photon momentum \( q^{\mu } \) and the average 
nucleon momentum \( P^{\mu } \) are collinear
along the \( z \)-axis and in opposite direction, one can parametrize
the non-perturbative object in the lower blobs of
Fig.~\ref{fig:handbags} as~:
\begin{eqnarray}
&& {{P^{+}}\over {2\pi }}\, \int dy^{-}e^{ixP^{+}y^{-}} 
\langle p^{'}|\bar{\psi }_{\beta }(-y/2) \psi _{\alpha}(y/2) 
|p\rangle {\Bigg |}_{y^{+}=\vec{y}_{\perp }=0} \nonumber \\
&=& {1\over 4}\left\{ ({\gamma ^{-}})_{\alpha \beta }
\left[ H^{q}(x,\xi ,t)\; \bar{N}(p^{'})\gamma ^{+}N(p)\, 
\right. \right.\nonumber\\
&&\hspace{2cm}\left. +\, E^{q}(x,\xi ,t)\; \bar{N}(p^{'})i\sigma ^{
+\kappa }{{\Delta _{\kappa }}\over {2m_{N}}}N(p)\right] \nonumber \\
&& \;+({\gamma _{5}\gamma ^{-}})_{\alpha \beta }
\left[ \tilde{H}^{q}(x,\xi ,t)\; \bar{N}(p^{'})\gamma ^{+}\gamma _{5}N(p)\, 
\right.\nonumber\\
&&\left. \left. \hspace{2cm}+\, \tilde{E}^{q}(x,\xi ,t)\; 
\bar{N}(p^{'})\gamma _{5}{{\Delta ^{+}}\over {2m_{N}}} N(p) \right]
\right\} , \;\;\;\;
\label{eq:qsplitting} 
\end{eqnarray}
where \( \psi  \) is the quark field, \( N \) the nucleon spinor 
and \( m_{N} \) the nucleon mass. 
The {\it lhs} of Eq.~(\ref{eq:qsplitting}) can be interpreted as a Fourier
integral along the light-cone distance $y^-$ of a quark-quark
correlation function, representing the process where 
a quark is taken out of the
initial nucleon (with momentum $p$) at the space-time point $y/2$, and
is put back in the final nucleon (with momentum $p'$) at the space-time
point $-y/2$. This process takes place at equal light-cone time ($y^+
= 0$) and at zero transverse separation ($\vec y_\perp = 0$) between
the quarks. The resulting one-dimensional Fourier integral along the
light-cone distance $y^-$ is with respect to the quark light-cone
momentum $x P^+$. 
The {\it rhs} of Eq.~(\ref{eq:qsplitting}) parametrizes this
non-perturbative object in terms of four SPD's, according to whether
they correspond to a vector operator $(\gamma^-)_{\alpha \beta}$ or
an axial-vector operator $(\gamma_5 \gamma^-)_{\alpha \beta}$ at the
quark level. The vector operator corresponds at the nucleon side
to a vector transition (parametrized by the function $H^q$, for a quark
of flavor $q$) and 
a tensor transition (parametrized by the function $E^q$). 
The axial-vector operator corresponds at the nucleon side
to an axial-vector transition (function $\tilde H^q$) 
and a pseudoscalar transition (function $\tilde E^q$).    
\newline
\indent
In Fig.~\ref{fig:handbags}, the variable $x$ runs from -1 to 1.
Therefore, the momentum fractions ($x + \xi$ or $x - \xi$) of the
active quarks can either be positive or negative. Since positive
(negative) momentum fractions correspond to quarks (antiquarks), it
has been noted in \cite{Rad96a} that in this way, one can  
identify two regions for the SPD's~: 
when $x > \xi$ both partons represent quarks, whereas for 
$x < - \xi$ both partons represent antiquarks. In these regions, 
the SPD's are the generalizations of the usual parton distributions from 
DIS. Actually, in the forward direction, the SPD's $H$ and $\tilde H$ 
reduce to the quark density distribution $q(x)$ and 
quark helicity distribution $\Delta q(x)$ respectively, obtained from DIS~:
 \begin{equation}
\label{eq:dislimit}
H^{q}(x,0,0)\, =\, q(x)\; ,\hspace {0.5cm}
\tilde{H}^{q}(x,0,0)\, =\, \Delta q(x)\; .
\end{equation}
The functions $E$ and $\tilde E$ are not measurable
through DIS because the associated tensors 
in Eq.~(\ref{eq:qsplitting}) vanish in the forward limit ($\Delta \to 0$). 
Therefore, $E$ and $\tilde E$ are new leading twist functions, which
are accessible through the  
hard exclusive electroproduction reactions, discussed in the following.  
\newline
\indent
In the region $ -\xi < x < \xi$, one parton connected to the lower
blob in Fig.~\ref{fig:handbags} represents a
quark and the other one an antiquark. In this region, the SPD's 
behave like a meson distribution amplitude and contain completely new
information about nucleon structure, because the region 
$ -\xi < x < \xi$ is absent in DIS, which corresponds to the limit 
$\xi \to 0$.   
\newline
\indent
Besides coinciding with the quark distributions at vanishing momentum
transfer, the skewed parton distributions have interesting links with other
nucleon structure quantities. The first moments of the SPD's are related to
the elastic form factors (FF) 
of the nucleon through model independent sum rules.  
By integrating Eq.~(\ref{eq:qsplitting}) over \( x \), one
obtains the following relations for one quark flavor : 
\begin{eqnarray}
\int_{-1}^{+1}dx\, H^{q}(x,\xi ,t)\, &=&\, F_{1}^{q}(t)\, , \nonumber\\
\int _{-1}^{+1}dx\, E^{q}(x,\xi ,t)\, &=&\, F_{2}^{q}(t)\, , \nonumber\\
\int_{-1}^{+1}dx\, \tilde{H}^{q}(x,\xi ,t)\, &=&\, g_{A}^{q}(t)\, , \nonumber\\
\int _{-1}^{+1}dx\,\tilde{E}^{q}(x,\xi ,t)\, &=&\, h_{A}^{q}(t)\,. 
\label{eq:ffsumrule} 
\end{eqnarray}
The elastic FF for one quark flavor on the {\it rhs} of 
Eqs.~(\ref{eq:ffsumrule}) are related 
to the physical ones (restricting oneself to \( u,d \) and \( s \) 
quark flavors) as~: 
\begin{equation}
\label{eq:vecff}
F_{1}^{u}\, =\, 2F_{1}^{p}+F_{1}^{n}+F_{1}^{s}\; ,  
\hspace {0.5cm}F_{1}^{d}\, =\, 2F_{1}^{n}+F_{1}^{p}+F_{1}^{s}\; ,
\end{equation}
where \( F_{1}^{p} \) and \( F_{1}^{n} \) are the usual proton and neutron 
Dirac FF respectively, and where \( F_{1}^{s} \) 
is the strangeness form factor. 
Relations similar to Eq.~(\ref{eq:vecff}) hold for the
Pauli FF \( F_{2}^{q} \). For the axial vector FF one uses 
the isospin decomposition~: 
\begin{equation}
\label{eq:axff}
g_{A}^{u}\, =\, {1\over 2}g_{A}+{1\over 2}g_{A}^{0}\; , 
\hspace {0.5cm}g_{A}^{d}\, =\, -{1\over 2}g_{A}+{1\over
  2}g_{A}^{0}\; , 
\end{equation}
where $g_A (g_A^0)$ are the isovector (isoscalar) axial FF
respectively. Similar relations exist for $h_A$. 
The isovector axial form factor \( g_{A} \) is known from experiment, with
\( g_{A}(0)\approx 1.267 \). The induced pseudoscalar form factor 
$h_A$ contains an important pion pole contribution, through the
partial conservation of the axial current (PCAC). 
\newline
\indent
A lot of the recent interest and activity in this field has been
triggered by the observation of \cite{Ji97} that the SPD's may shed a
new light on the ``spin-puzzle''.
Starting from a (color) gauge-invariant decomposition of the nucleon spin~: 
$1/2\, =\, J_{q}\, +\, J_{g}\,$ ,
where \( J_{q} \) and \( J_{g} \) are the total quark and gluon
angular momentum respectively, it was shown in \cite{Ji97} that 
the second moment of the unpolarized SPD's at \( t=0 \) gives 
\begin{equation}
\label{eq:dvcsspin}
\hspace{-0.2cm} J_{q}\, = {1\over 2} \int _{-1}^{+1}dx\, x\, 
\left[ H^{q}(x,\xi ,t=0)+E^{q}(x,\xi ,t=0)\right] ,
\end{equation}
 and this relation is independent of \( \xi  \). 
The quark angular momentum \( J_{q} \) decomposes as : 
$J_{q}=\Delta \Sigma/2 + L_{q}$ ,
where \( \Delta \Sigma /2 \) and \( L_{q} \) are the quark spin
and orbital angular momentum respectively. As \( \Delta \Sigma  \) is measured
through polarized DIS experiments, 
a measurement of the sum rule of Eq.~(\ref{eq:dvcsspin}) in
terms of the SPD's, provides a model independent way to determine the quark
orbital contribution $L_q$ to the nucleon spin.
\newline
\indent
Ultimately, one wants to extract the SPD's from data, but in order to evaluate 
the electroproduction observables, and to study their sensitivity to the new
physics, one needs an educated guess for the SPD's. 
In Ref.~\cite{Vdh99}, a model for the SPD's was constructed using a 
$\xi$-dependent product ansatz (for the double distributions 
introduced in Ref.~\cite{Rad99}) of a quark distribution 
and an asymptotic ``meson-like'' distribution amplitude 
(see Ref.~\cite{Vdh99} for more details). For the quark
distributions, the MRST98 parametrization \cite{MRST98} is used as input.  
The $t$-dependence of the model for the SPD's is given by the corresponding 
FF (Dirac form factor for $H$, axial form factor for $\tilde H$), 
so that the first moments of the
SPD's are satisfied by construction. As an example, 
the $d$-quark SPD (formerly also denoted as off-forward parton
distribution (OFPD)), using the above described ansatz, 
is shown in Fig.~\ref{fig:ofpdxi}.  

One observes from Fig.~\ref{fig:ofpdxi} the transition from a quark
distribution ($\xi = 0$) to a meson distribution amplitude ($\xi =
1$). Model calculation of the SPD's are currently possible within the
QCD chiral models for intermediate $x_B$. In particular,
a calculation \cite{Pet98} in the chiral quark soliton model 
(see Ref.~\cite{Chr96} for a review) 
found a strong dependence of the SPD's on $\xi$ and fast 
``crossovers'' at $|x| = \xi$. Such behavior is related to the fact 
that the SPD's in the region $-\xi < x < \xi$ have properties of meson
distribution amplitudes. In particular for the SPD $H$, this can be
seen as being due to a scalar-isoscalar two-pion exchange
contribution \cite{Pol99}, 
indicating that the SPD's are qualitatively a richer source of
nucleon structure information than ordinary parton distributions. 
One may expect that eventually it will be possible to calculate SPD's
for intermediate $x_B$ using lattice QCD.

\subsection{Leading order amplitudes and observables for 
DVCS and hard meson electroproduction}

\begin{figure}[h]
\epsfxsize=7.cm
\vspace{-2.3cm}
\centerline{\hspace{1.2cm}\epsffile{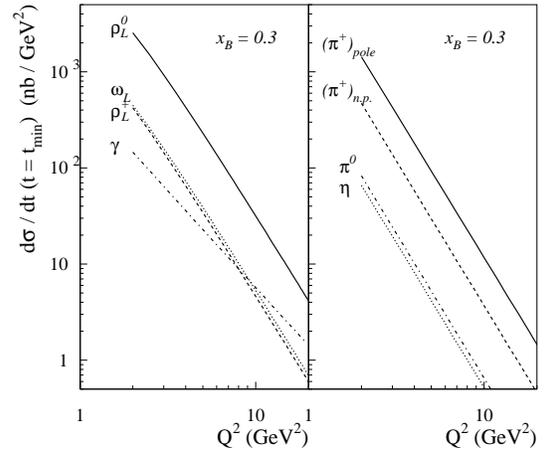}}
\vspace{-1.9cm}
\caption[]{\small Scaling behavior of the L.O. predictions
  for the forward differential electroproduction cross section 
on the proton, for vector mesons (left panel) 
and pseudoscalar mesons (right panel), as calculated in Ref.~\cite{Vdh99}. 
For the $\pi^+$ channel, the pion pole contribution
(full line, $(\pi^+)_{pole}$) is shown separately from the 
$\tilde H$ contribution (dashed line, $(\pi^+)_{n.p.}$). 
Also shown is the scaling behavior
of the forward transverse cross section $d \sigma_T/d t$ 
for the leading order DVCS cross section (dashed-dotted line in left panel).}
\label{fig:scaling}
\end{figure}

The leading order (L.O.) DVCS amplitude in the forward direction 
 is given \cite{Ji97} by the handbag diagram  
shown on the left panel of Fig.~\ref{fig:handbags} (the crossed
diagram which is not shown is also understood). 
A formal factorization proof for DVCS has been given in 
Refs.~\cite{Ji98a,Col99}.
\newline
\indent
To calculate the DVCS amplitude in the Bjorken regime, it is 
natural to express the momenta in the process 
($q^\mu$ of the virtual photon, $q'^{\mu}$
 of the real photon, and $P^\mu$ denoting the average nucleon momentum) 
in terms of the lightlike vectors  
\begin{equation}
\tilde p^\mu = {{P^+} \over {\sqrt{2}}} (1,0,0,1)\;, \hspace{0.5cm}
n^\mu = {1 \over {P^+\sqrt{2}}} (1,0,0,-1) \;.
\end{equation}
Using the parametrization of Eq.~(\ref{eq:qsplitting}) 
for the bilocal quark operator, 
the L.O. DVCS tensor $H^{\mu \nu}_{L.O. \, DVCS} $ 
(defined e.g. in \cite{Gui98}) follows from the handbag diagrams as~: 
\begin{eqnarray}
&&H^{\mu \nu }_{L.O.\, DVCS}\nonumber \\
&=& {1\over 2}\, \left[ \tilde{p}^{\mu }n^{\nu }+\tilde{p}^{\nu}n^{\mu }
- g^{\mu \nu }\right] \; \nonumber\\
&& \times \int _{-1}^{+1}dx \left[ {1\over {x-\xi +i\epsilon }} 
+{1\over {x+\xi -i\epsilon}}\right] \nonumber \\
&& \times \left[ H^{p}_{DVCS}(x,\xi ,t)\; \bar{N}(p^{'})\gamma 
.n N(p) \right. \nonumber\\ 
&&\left. \hspace{0.3cm} +\, E^{p}_{DVCS}(x,\xi ,t)\; 
\bar{N}(p^{'})i\sigma ^{\kappa \lambda } 
{{n_{\kappa }\Delta _{\lambda }}\over {2m_{N}}}N(p)\right] \nonumber \\
&+& \; {1\over 2}\, \left[ -i\varepsilon ^{\mu \nu \kappa \lambda } 
\tilde{p}_{\kappa }n_{\lambda }\right] \; 
\int _{-1}^{+1}dx\left[ {1\over {x-\xi +i\epsilon }}
-{1\over {x+\xi -i\epsilon }}\right] \nonumber \\
&& \times \left[ \tilde{H}^{p}_{DVCS}(x,\xi ,t)  
\bar{N}(p^{'})\gamma .n\gamma _{5}N(p) \right.\nonumber\\
&&\left. \hspace{0.3cm}+\,\tilde{E}^{p}_{DVCS}(x,\xi ,t)
\bar{N}(p^{'})\gamma_{5}{{\Delta \cdot n}\over {2m_{N}}}N(p)\right] .\;
\label{eq:dvcsampl} 
\end{eqnarray}
On the {\it rhs} of the DVCS tensor
of Eq.~(\ref{eq:dvcsampl}), the SPD's $H, \tilde H, E, \tilde E$ enter
in a convolution integral over the quark momentum fraction $x$. 
This is a qualitative difference compared with the DIS amplitude, 
where one is only sensitive (through the optical theorem) 
to the imaginary part of the forward double virtual Compton amplitude.  
We refer to Ref.~\cite{Gui98} for details and for the
formalism to calculate DVCS observables starting from the DVCS tensor of 
Eq.~(\ref{eq:dvcsampl}). 
\newline
\indent
The leading order DVCS amplitude corresponding to 
Eq.~(\ref{eq:dvcsampl}), is exactly
gauge invariant with respect to the virtual photon, i.e. 
\( q_{\nu }\, H^{\mu \nu }_{L.O.\, DVCS}=0 \).
However, electromagnetic gauge invariance is violated by the 
real photon except in the forward direction. 
This violation of gauge invariance is a higher twist (twist-3) effect 
compared to the leading order term $ H^{\mu \nu }_{L.O.\, DVCS} $.
Since $q^{'}_{\mu }\, H^{\mu \nu }_{L.O.DVCS}\sim \Delta _{\perp }$, 
an improved DVCS amplitude linear in $\Delta _{\perp }$ has been
proposed in Ref.~\cite{Gui98} to restore gauge invariance (in the
nonforward direction) in a heuristic way~:
\begin{equation}
\label{eq:dvcsgaugeinv}
H^{\mu \nu }_{DVCS}\, =\, H^{\mu \nu }_{L.O.\, DVCS}\, +\,
{{\tilde{p}^{\mu }}\over {\left( \tilde{p}\cdot q^{'}\right) }}\; 
\left( \Delta _{\perp }\right) _{\lambda }\, H^{\lambda \nu }_{L.O.\, DVCS}\; ,
\end{equation}
leading to a correction term to the L.O. DVCS amplitude of order 
$O\left( \Delta_{\perp} / Q \right)$.  
\newline
\indent
Very recently, the gauge invariance of the DVCS amplitude was
addressed in much more detail in several works
\cite{Ani00,Pen00,Bel00b}. It was found that the twist-3 terms which
restore gauge invariance (to twist-4 accuracy) involve two
contributions. First there are terms proportional to the twist-2 SPD's
of Eq.~(\ref{eq:dvcsampl}), which were found to completely
coincide with the improved DVCS amplitude 
of Eq.~(\ref{eq:dvcsgaugeinv}). In addition, there are
terms which are characterized by new `transverse'
SPD's (see Refs.~\cite{Ani00,Pen00,Bel00b} for details). These latter
functions are suppressed by one power $1/Q$ compared with the
contribution of the twist-2 SPD's in DVCS cross sections, and could in
principle be separated by measuring DVCS observables over a 
sufficiently large $Q^2$ range (see e.g. \cite{Die97} for tests of the handbag
approximation to DVCS). 
In view of current DVCS experiments which are performed or planned at
$Q^2$ in the few GeV$^2$ range only, the numerical importance of those
additional contributions remains to be investigated. 
\newline
\indent
Besides the DVCS process, a factorization proof was also given 
for the L.O. meson electroproduction amplitudes in the
Bjorken regime \cite{Col97,Rad96b}, which is illustrated on the right panel
of Fig.~\ref{fig:handbags}. This factorization theorem only applies
when the virtual photon is {\it longitudinally} polarized. 
In the valence region, the L.O. amplitude 
\({\mathcal{M}}^{L} \) for meson production by a longitudinal
photon consists of evaluating Fig.~\ref{fig:handbags} (right panel,
where only one of the four L.O. diagrams is shown) 
with the one-gluon exchange diagrams as hard scattering kernel. 
In this way, the L.O. expressions for  
\( \rho ^{0}_{L} \) (longitudinally polarized vector meson) and 
\( \pi ^{0} \) electroproduction were calculated in \cite{Vdh98} 
(see also Ref.~\cite{Man98}) as~: 
\begin{eqnarray}
{\mathcal{M}}^{L}_{\rho ^{0}_{L}}\;&=&\; -ie\, {4\over 9}\, 
{1\over {Q}}\; \left[ \, \int _{0}^{1}dz{{\Phi _{\rho }(z)}\over z}\right]  
\nonumber\\
&\times& {1\over 2}\, \int _{-1}^{+1}dx\left[ {1\over {x-\xi +i\epsilon }}  
+{1\over {x+\xi -i\epsilon }}\right] \nonumber \\
&\times& (4\pi \alpha _{s})
\left\{ H^{p}_{\rho ^{0}_{L}}(x,\xi ,t) 
\bar{N}(p^{'})\gamma .n N(p) \right. \nonumber\\
&&\left. \hspace{1cm}+E^{p}_{\rho ^{0}_{L}}(x,\xi ,t)\bar{N}(p^{'})
i\sigma ^{\kappa \lambda }{{n_{\kappa }\Delta _{\lambda }}\over 
{2m_{N}}}N(p)\right\}, \;\;\;\; \label{eq:rhoampl} 
\end{eqnarray}
\begin{eqnarray}
{\mathcal{M}}^{L}_{\pi^{0}}\;&=&\; 
-ie\, {4 \over 9}\, {1 \over Q}\,  
\left[ \int _{0}^{1}dz {{\Phi_{\pi}(z)}\over z}\right] \,  \nonumber\\ 
&\times& {1\over 2}\, \int _{-1}^{+1}dx\left[ {1\over {x-\xi +i\epsilon }}  
+{1\over {x+\xi -i\epsilon }}\right] \nonumber\\
&\times& (4\pi \alpha _{s}) 
\left\{ \tilde{H}^{p}_{\pi ^{0}}(x,\xi ,t) 
\bar{N}(p^{'})\gamma .n\gamma _{5} N(p) \right. \nonumber\\
&&\left. \hspace{1cm}+\tilde{E}^{p}_{\pi ^{0}}(x,\xi,t)  
\bar{N}(p^{'})\gamma_{5} {{\Delta \cdot n}\over {2m_N}} N(p)\right\}, \;\;
\label{eq:piampl} 
\end{eqnarray}
where $\alpha_s$ is the QCD coupling constant. Because 
the quark helicity is conserved in the hard scattering process, 
one finds the interesting result that
the vector meson electroproduction amplitude depends only on the {\it
  unpolarized} SPD's $H$ and $E$, whereas the pseudoscalar meson
electroproduction amplitudes depend only on the {\it polarized}
SPD's $\tilde H$ and $\tilde E$. In contrast, the DVCS amplitude of 
Eq.~(\ref{eq:dvcsampl}) depends on both the unpolarized and polarized
SPD's. Another difference from DVCS, is the fact that
the meson electroproduction amplitudes require 
additional non-perturbative input from the 
meson distribution amplitudes  
\( \Phi _{\rho }(z) \) and \( \Phi _{\pi }(z) \) respectively, 
for which the asymptotic forms are taken in the calculations. 
From Eqs.~(\ref{eq:rhoampl},\ref{eq:piampl}), 
one furthermore sees that the L.O. longitudinal amplitudes for
meson electroproduction behave as \( 1/Q \). 
At large \( Q^{2} \), fixed \( x_{B} \) and fixed \(t \), 
this leads to a \( 1/Q^{6} \) behavior for the longitudinal cross 
section \( d\sigma _{L}/dt \), which provides an experimental
signature (scaling) of the leading order mechanism.   
Expressions analogous to Eqs.~(\ref{eq:rhoampl}, \ref{eq:piampl}) have
also been worked out for the charged meson channels $\rho^\pm,
\pi^\pm$ as well as for the $\omega, \phi$ and $\eta$ channels 
(see Refs.~\cite{Fra99,Man99a,Man99b,Vdh99} for details).
\newline
\indent 
According to the considered reaction, the SPD's 
enter in different combinations due to the  charges and isospin factors. 
For DVCS on the proton, the combination is  
\begin{equation}
H^p_{DVCS}(x,\xi,t) \,=\,{4 \over 9} H^{u} \,+\, {1 \over 9} H^{d} 
\,+\,{1 \over 9} H^{s} \;,
\end{equation}
and similarly for $\tilde H$, $E$ and $\tilde E$. For  
electroproduction of $\rho^0$ and $\pi^0$ on the proton, 
the isospin structure yields the combination 
\begin{eqnarray}
H^p_{\rho^0}(x,\xi,t) &=& {1 \over {\sqrt{2}}} \left\{ {2 \over 3} H^{u} 
+ {1 \over 3} H^{d} \right\} , \\
\tilde H^p_{\pi^0}(x,\xi,t) &=&{1 \over {\sqrt{2}}} \left\{
{2 \over 3} \tilde H^{u} + {1 \over 3} \tilde H^{d} \right\} ,
\end{eqnarray}
and similar for $E$ and $\tilde E$. Corresponding relations for 
the $\rho^\pm, \omega, \phi, \pi^\pm$ and $\eta$ channels can be found in
Refs.~\cite{Fra99,Man99a,Man99b,Vdh99}. 
Therefore, the measurements of the different meson
production channels are sensitive to different combinations of the same
universal SPD's, and allow us to perform a flavor separation of
the SPD's, provided one is able to deconvolute the SPD's from the
leading order amplitudes.  
\newline
\indent
Some representative results for DVCS and meson
electroproduction observables using the $\xi$-dependent 
ansatz for the SPD's, are shown in the following. 
More detailed results can be found in Refs.~\cite{Vdh98,Gui98,Vdh99}.
\newline
\indent 
Before considering the extraction of the SPD's from electroproduction
data, it is compulsory to demonstrate that the scaling regime has been reached.
In Fig.~\ref{fig:scaling}, the forward longitudinal 
electroproduction cross sections are shown 
as a function of \( Q^{2} \) and the L.O. predictions are compared 
for different mesons. The L.O. amplitude 
for longitudinal electroproduction of mesons was
seen to behave as \( 1/Q \), leading to a 
\( 1/Q^{6} \) scaling behavior for \( d\sigma _{L}/dt \).  

\begin{figure}[h]
\epsfxsize=7.75 cm
\vspace{-1.25cm}
\centerline{\epsffile{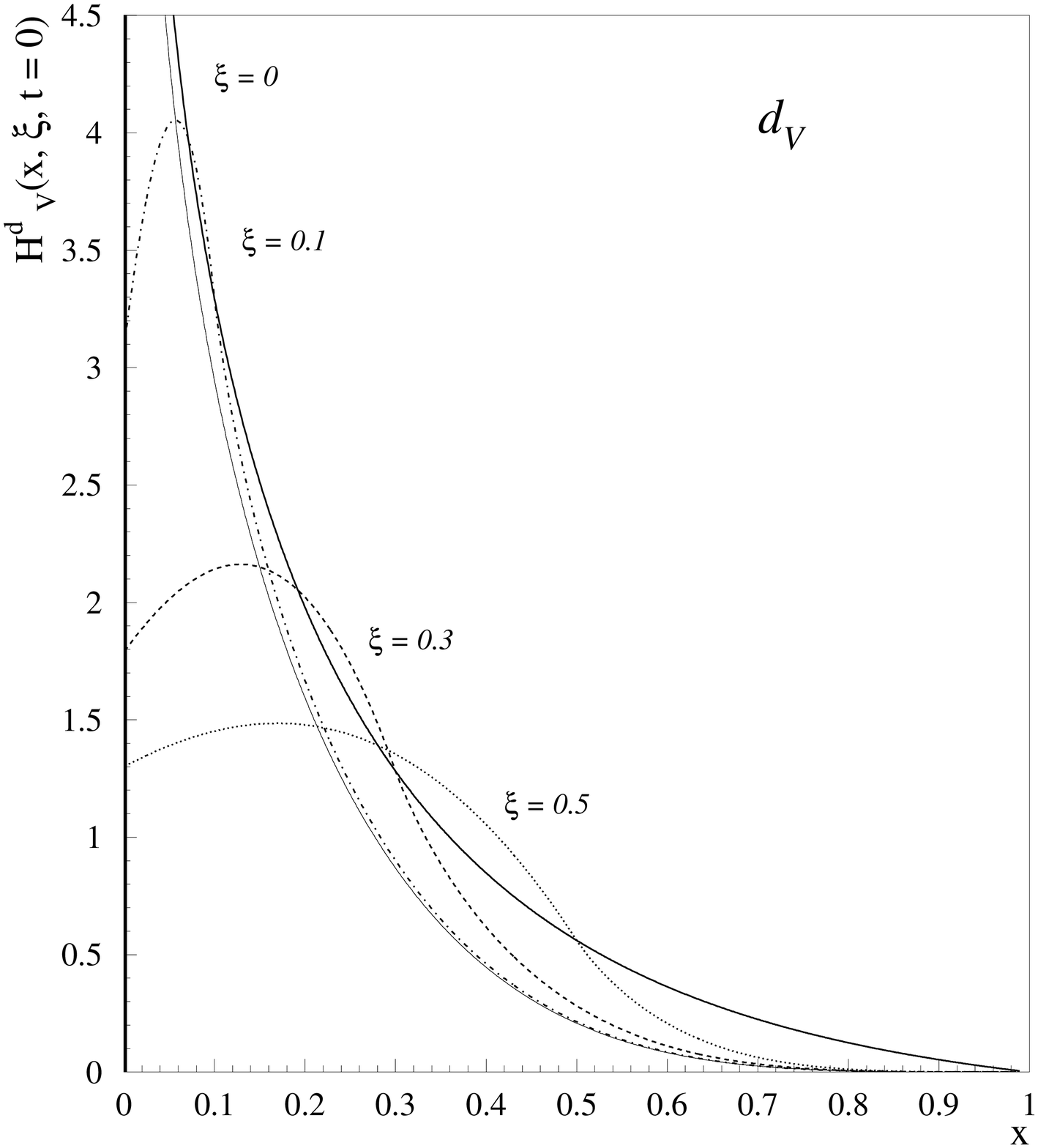}}
\caption[]{\small Valence $d$-quark contribution to the SPD $H^d$ 
at $t = 0$ for different values of $\xi$ as indicated on the curves,
calculated with the ansatz as in Fig.~\ref{fig:ofpdxi}.
The thin (lower) solid curve ($\xi = 0$) corresponds to the ordinary
$d$-quark distributions, whereas the thick (upper) solid curve to the
envelope function $H_V^d(x = \xi, \xi, t = 0)$ as measured through the
SSA.}
\label{fig:down_xiscan}
\epsfxsize=8.7 cm
\vspace{0.2cm}
\centerline{\epsffile{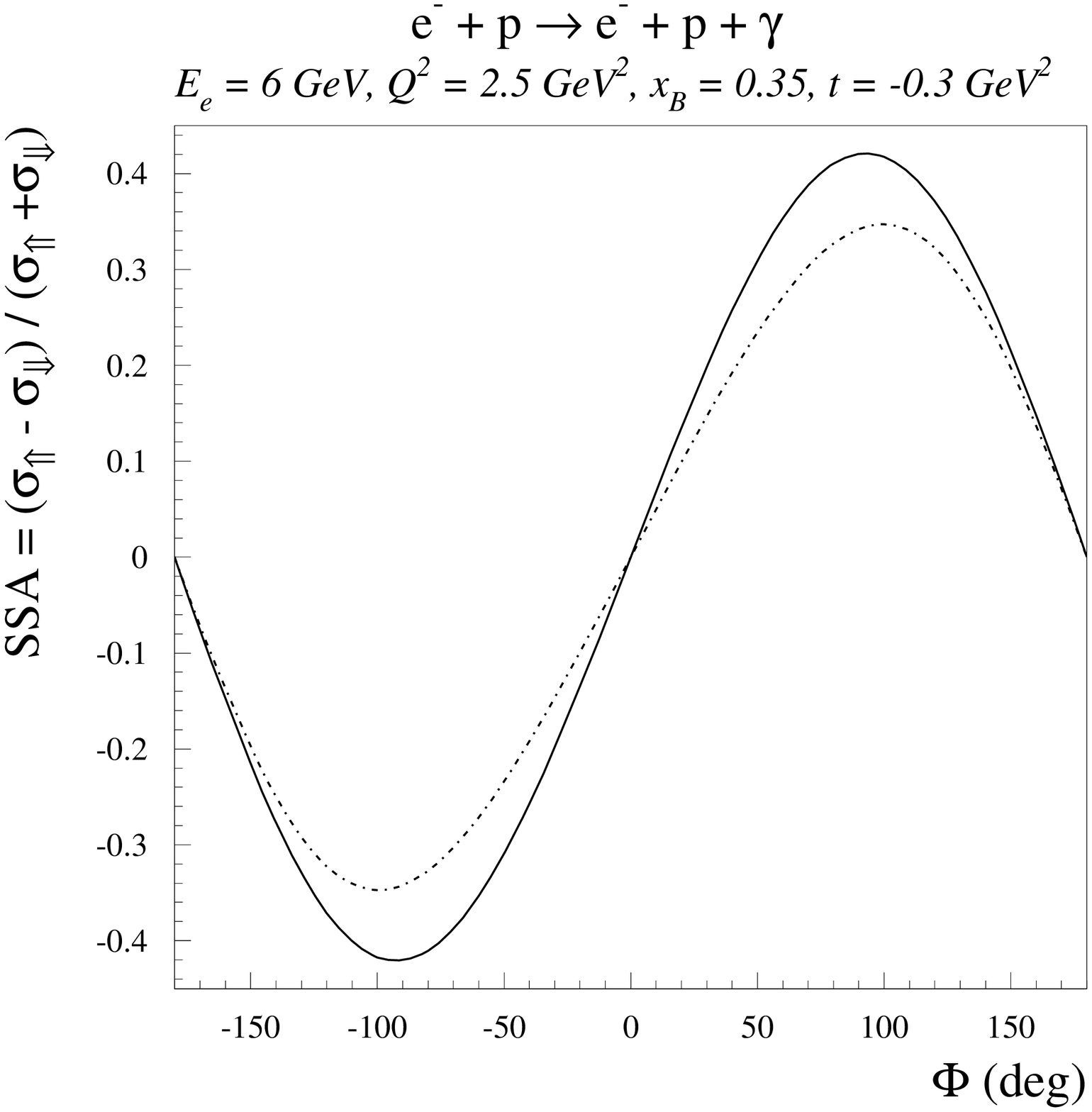}}
\vspace{-.4cm}
\caption[]{\small Single spin asymmetry for DVCS in JLab kinematics. 
A comparison is shown of the asymmetries for a $\xi$-independent
ansatz for the SPD's as in Ref.~\cite{Gui98} (dashed-dotted curves), 
and a $\xi$-dependent ansatz for the SPD's as 
in Fig.~\ref{fig:ofpdxi} (full curve).}
\label{fig:ssa_jlab}
\end{figure}

By comparing the different vector meson channels in Fig.~\ref{fig:scaling},
one sees that the \( \rho ^{0}_{L} \) channel yields the largest cross section.
The \( \omega _{L} \) channel in the valence region (\( x_{B}\approx  \) 0.3)
is about a factor of 5 smaller than the \( \rho ^{0}_{L} \) channel, which
is to be compared with the ratio at small \( x_{B} \) (in the
diffractive regime) 
where \( \rho ^{0} \) : \( \omega  \) = 9 : 1. The \( \rho _{L}^{+} \)
channel, which is sensitive to the isovector combination of the unpolarized
SPD's, yields a cross section comparable to the \( \omega _{L} \)
channel. The \( \rho _{L}^{+} \) channel is interesting as there is no
competing diffractive contribution, 
and therefore allows to test directly the quark SPD's.
The three vector meson channels (\( \rho _{L}^{0} \), \( \rho _{L}^{+} \),
\( \omega _{L} \)) are highly complementary in order to perform a
flavor separation of the unpolarized SPD's \( H^{u} \) and \( H^{d} \). 
\newline
\indent
A dedicated experiment is planned at JLab at 6 GeV 
in the near future \cite{Guid98} 
to investigate the onset of the scaling behavior for $\rho^0_L$
electroproduction in the valence region 
($Q^2 \approx 3.5$ GeV$^2$, $x_B \approx 0.3$). 
\newline
\indent
For the pseudoscalar mesons which involve 
the polarized SPD's, one remarks in Fig.~\ref{fig:scaling} 
the prominent contribution 
of the charged pion pole to the \( \pi ^{+} \) cross section. 
For the  contribution proportional to the SPD \( \tilde{H} \), 
it is also seen that the \( \pi ^{0} \)
channel is about a factor of 5 below the \( \pi ^{+} \) channel due to
isospin factors. In the \( \pi ^{0} \) channel,
the \( u \)- and \( d \)-quark polarized SPD's enter with the same sign,
whereas in the \( \pi ^{+} \) channel, they enter with opposite signs. As the
polarized SPD's are constructed here from the corresponding polarized parton
distributions, the difference between the 
predictions for the \( \pi ^{0} \) and
\( \pi ^{+} \) channels results from the fact that the polarized \( d \)-quark
distribution is opposite in sign to the polarized \( u \)-quark
distribution. For the $\eta$ channel, the ansatz for the SPD $\tilde
H$ based on the polarized quark distributions yields a prediction comparable
to the $\pi^0$ cross section. 
\newline
\indent
For the \( \gamma  \) leptoproduction in the few GeV beam energy range, 
the cross section is dominated by the Bethe-Heitler (BH)
process (see Ref.~\cite{Gui98}). 
However, it was suggested in Ref.~\cite{Gui98} that an exploration of 
DVCS might be possible if the beam is  polarized.
The electron single spin asymmetry (SSA) does not vanish out of plane 
due to the interference between the purely real BH process and the imaginary
part of the DVCS amplitude. 
Because the SSA measures the imaginary part of the DVCS amplitude, it
is directly proportional to a linear combination of the SPD's along
the line $x = \xi$. In fact, the SSA maps out an `envelope' function, 
e.g. $H(x = \xi, \xi, t)$, as shown e.g. in Fig.~\ref{fig:down_xiscan} for
the valence down quark SPD in the ansatz corresponding with 
Fig.~\ref{fig:ofpdxi}. 
\newline
\indent
In Fig.~\ref{fig:ssa_jlab}, it is shown that the SSA yields a sizeable
asymmetry for JLab kinematics, and displays a sensitivity to the 
$\xi$-dependent shape of the SPD's. 
An experiment to measure the SSA for DVCS has very recently been 
proposed at JLab at 6 GeV \cite{Ber99}. 
The SSA for DVCS is at present also measured at HERMES \cite{Arm99}. 
\newline
\indent
Going up in energy, the increasing virtual photon flux factor boosts the 
DVCS part of the $\gamma$ leptoproduction cross section, making it
more important compared to the BH contribution. 
This provides a nice opportunity for COMPASS at 200 GeV beam energy, 
where experiments have been proposed for DVCS \cite{d'Ho99}
and meson electroproduction \cite{Poc99}.

\subsection{Extension to hard exclusive electroproduction of decuplet
  baryons}

\begin{figure}[h]
\vspace{-0.65cm}
\epsfxsize=7.5 cm
\epsfysize=9.5 cm
\centerline{\epsffile{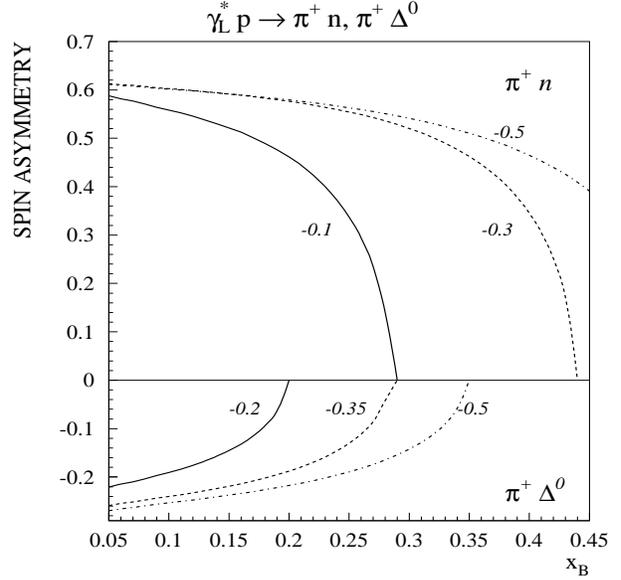}}
\vspace{-1.3cm}
\caption[]{\small Transverse spin asymmetries  
  for the longitudinal electroproduction of 
$\pi^+ n$ and $\pi^+ \Delta^0$, 
at different values of $t$ (in (GeV/c)$^2$). Figure from Ref.~\cite{Fra00}.}
\label{fig:pidelasymm}
\end{figure}

In the previous section, the main focus  
were the reactions $\gamma^* + N \to \gamma + N'$
and $\gamma^*_L + N \to M + N'$ with $M$ a meson, and where $N'$ is an
octet baryon. One of the intriguing questions of medium-energy QCD dynamics is
the differences and similarities in the structure of  baryons belonging 
to the different $SU(3)_f $ multiplets. 
In particular, a naive constituent quark model predicts that they are
similar, while there are suggestions that due to a strong attraction
between the quarks in the spin-isospin zero channel, diquark 
correlations should be important in the baryon octet but not in the
decuplet \cite{Sch98}.
At the same time the chiral models suggest that in the 
limit of a large number of colors (large $N_c$) of QCD, which is
known to be a useful guideline, nucleons and
$\Delta$ isobars are different rotational excitations 
of the same soliton \cite{Adk83,Dia88}.
\newline
\indent
For these studies, the potential of the process 
$\gamma^*_L +N \to \pi +\Delta $ as well as the DVCS process 
$\gamma^* +N \to \gamma +\Delta $, was explored in Ref.~\cite{Fra00}. 
In addition, the study of the processes with production of
decuplet baryons has also a practical usefulness, because 
in the experiments with low resolution 
in the mass of the recoiling system ($\Delta M \approx$ 300 MeV for
HERMES in the current set-up),  
the estimates of $\Delta$ production are necessary
to extract the $N \to N$ SPD's from such data.
\newline
\indent
In Ref.~\cite{Fra00}, a new set of SPD's were introduced 
for the axial $N\to \Delta$ (isovector) 
transition, denoted as $C_i^{(3)}$, which enter into 
$\pi \Delta$ electroproduction~:
\begin{eqnarray}
\hspace{-0.1cm} &&{{P^+} \over {2 \pi}} \int d y^{-} e^{i x P^{+} y^{-}} 
\langle \Delta^{+}|\bar \psi(-y/2) {\Dirac n} \gamma^5 
\psi(y/2) |N \rangle {\Bigg |}_{y^{+}=\vec{y}_{\perp }=0} 
\nonumber \\
&&= \bar \psi^\beta(p') \; \left[\,
C_1^{(3)}\left(x, \xi, t \right) n_\beta \right. \nonumber\\
&&\left. \hspace{1.5cm}+\; 
C_2^{(3)}\left(x, \xi, t \right) \frac{\Delta_\beta(n\cdot\Delta)}{m_N^2}
+\ldots \right] \; N(p) ,
\label{eq:axialndelta}
\end{eqnarray}
where the same notations are used as before, and 
where $\psi^\beta(p')$ is the Rarita-Schwinger spinor for the 
$\Delta$ isobar. 
In Eq.~(\ref{eq:axialndelta}), the ellipses
denote other contributions which are suppressed at large $N_c$. 
For the $N \to \Delta$ DVCS process, besides the axial SPD's, also
vector SPD's enter which were also defined in Ref.~\cite{Fra00}. 
\newline
\indent
The observation that in the large $N_c$ limit, the nucleon
and $\Delta$ are rotational excitations of {\em the same} classical 
soliton object, allows us to derive a number of relations between
$N\to N$ and $N\to\Delta$ SPD's. For $C_1^{(3)}$ and $C_2^{(3)}$, 
these have the form \cite{Fra99}~:
\begin{eqnarray}
\label{eq:spdndelta}
C_1^{(3)}(x, \xi, t) &=& \sqrt{3}\, 
\widetilde H^{(3)}(x, \xi, t) , \\
C_2^{(3)}(x, \xi, t) &=& \sqrt{3}/4\, 
\widetilde E^{(3)}(x, \xi, t) \, ,
\end{eqnarray}
in terms of the (isovector) $N \to N$ 
SPD $\widetilde H^{(3)} = \widetilde H^u - \widetilde H^d$, 
and analogously for $\widetilde E^{(3)}$.           
\newline
\indent
Using the large $N_c$ relations of Eq.~(\ref{eq:spdndelta}), one can easily
derive the relations between the different cross sections for charged
pion production as 
$\sigma_L^{\gamma^* p \to \pi^+ n}:\sigma_L^{\gamma^* p \to \pi^+\Delta^0}:\sigma_L^{\gamma^* p \to \pi^-\Delta^{++}}:\sigma_L^{\gamma^* n \to \pi^- p}
\approx 1:0.5:1.25:0.8$.
\newline
\indent
Besides the cross section $\sigma_L$, the second
observable involving only longitudinal amplitudes and being 
a leading order observable for hard exclusive
meson electroproduction, is the single spin asymmetry,  
for a proton target polarized perpendicular to the reaction
plane (or the equivalent recoil polarization observable) \cite{Fra99}. 
These transverse spin asymmetries for 
$\pi^+ n$ and $\pi^+ \Delta^0$ are shown in Fig.~\ref{fig:pidelasymm}.  

It is obvious from Fig.~\ref{fig:pidelasymm}, that large transverse spin
asymmetries are predicted for these processes, 
related to the peculiar feature of chiral QCD.
As a consequence, investigations 
of these processes can provide unique tests of the
soliton type  approach to baryon structure. 
The spin asymmetries are likely to be less
sensitive to higher twist effects and hence can be explored
already using the HERMES detector and JLab at higher energies.
Furthermore, the study of these processes would allow one to make a more
reliable separation of the $\pi$ pole contribution in the
electroproduction of pions, which is mandatory for the measurement of
the pion elastic form factor at higher $Q^2$. 

\subsection{Power corrections to the leading order amplitudes}

\begin{figure}[ht]
\vspace{-.6cm}
\epsfxsize=7 cm
\epsfysize=9cm
\centerline{\hspace{.25cm} \epsffile{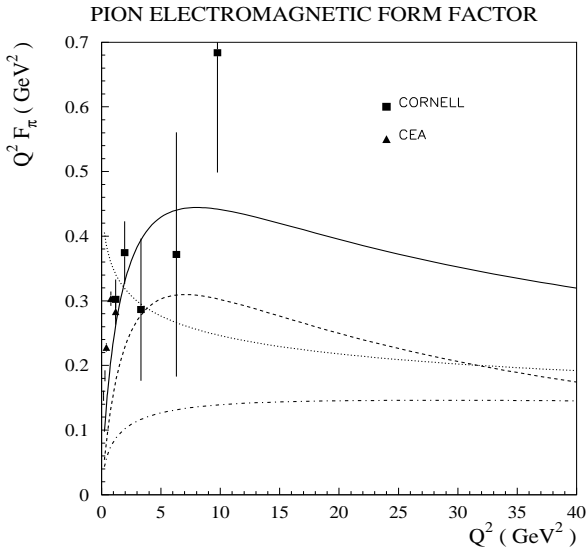}}
\vspace{-1.2cm}
\caption[]{\small Results for the $\pi$ electromagnetic form factor of
the leading order PQCD prediction without (dotted line) and with 
(dashed-dotted line ) inclusion of the corrections due to 
intrinsic transverse momentum dependence. 
The dashed curve shows the result for the soft overlap contribution 
and the total result (full line) 
is the sum of the dashed and dashed-dotted lines. 
Figure from Ref.~\cite{Vdh99}, where the
references of the data can also be found.}
\label{fig:piemff}
\end{figure}

When measuring hard electroproduction reactions in the region 
\( Q^{2}\approx 1-20 \) GeV\( ^{2} \), there arises the question of
the importance of power corrections to the leading order amplitudes, 
i.e. how fast does one approach the scaling regime predicted by the 
L.O. amplitudes. 
One source of power corrections is evident from the structure of the
matrix element of Eq.~(\ref{eq:qsplitting}) which defines the SPD's,
where the quarks are taken at zero transverse separation. This amounts
to neglect, at leading order, the quark's transverse
momentum compared with its large longitudinal (+ component) momentum. 
A first estimate of these corrections due to the quark's intrinsic transverse
momentum has been obtained in Ref.~\cite{Vdh99}, 
which is referred to for details.  
This correction is known to be important for a successful
description at the lower $Q^2$ values 
of the \( \pi ^{0}\gamma ^{*}\gamma  \) transition FF,
for which data exist in the range \( Q^{2}\approx 1-10 \) GeV\( ^{2} \). For
the pion elastic FF in the transition region before asymptotia
is reached, the power corrections due to both the transverse momentum 
dependence and the soft overlap mechanism 
(i.e. the process which does not proceed through one-gluon exchange) 
are quantitatively important. 
The result for the pion elastic FF is shown in 
Fig.~\ref{fig:piemff}, where it is seen that the
leading order PQCD result is approached only at very large \( Q^{2} \). The
correction including the transverse momentum dependence gives a substantial
suppression at lower \( Q^{2} \) (about a factor of two around 
\( Q^{2}\approx  \) 5 GeV\( ^{2} \)). 
At these lower \( Q^{2} \) values, the inclusion of the
transverse momentum dependence renders the PQCD calculation internally
consistent. 

\begin{figure}[h]
\vspace{-0.1cm}
\epsfysize=10.5 cm
\centerline{\hspace{.25cm} \epsffile{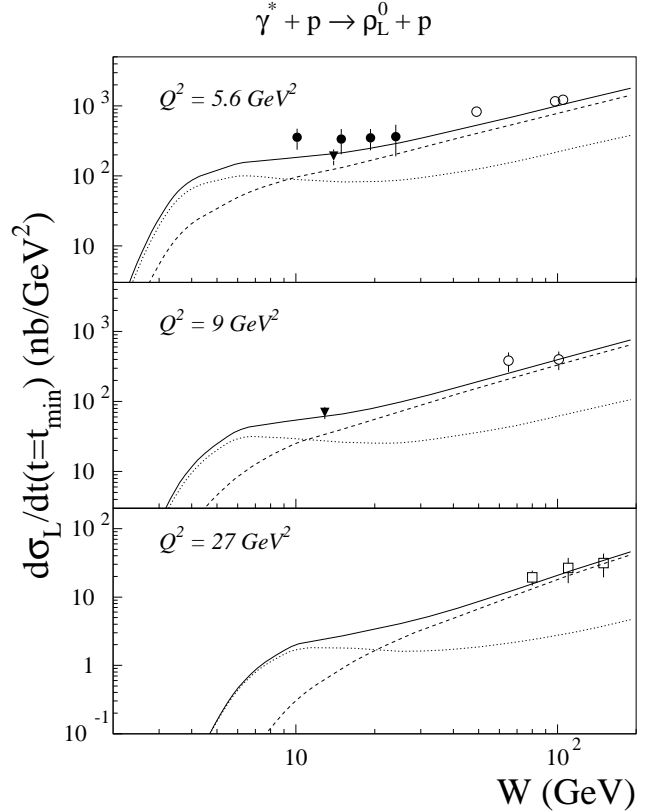}}
\vspace{-0.2cm}
\caption[]{\small Longitudinal forward differential cross section for 
$\rho^0_L$ electroproduction. Calculations compare the quark exchange
mechanism (dotted lines) with the two-gluon
exchange mechanism (dashed lines) and the sum of both (full
lines). Both calculations include the corrections due to intrinsic
transverse momentum dependence. 
The data are from NMC (triangles), 
E665 (solid circles), ZEUS 93 (open circles)  
and ZEUS 95 (open squares). Figure from Ref.~\cite{Vdh99}, where the
references of the data can also be found.}
\label{fig:rhotot2}
\end{figure}

These form factors were taken 
as a guide in Ref.~\cite{Vdh99} to estimate the corrections due to
the parton's intrinsic transverse momentum dependence in the DVCS and
hard meson electroproduction amplitudes. 
\newline
\indent
Although experimental data for \( \rho ^{0}_{L} \) electroproduction
at larger \( Q^{2} \) do not yet exist 
in the valence region (\( x_{B}\approx  \)
0.3), the reaction \( \gamma^{*}_L\, p\longrightarrow \rho ^{0}_{L}\, p \)
has been measured at smaller values of \( x_{B} \). 
Therefore, in Fig.~\ref{fig:rhotot2} the calculations are compared 
to those data, in order to study 
how the valence region is approached, in which one is sensitive
to the quark SPD's. For the purpose of this discussion, we call the mechanism
proceeding through the quark SPD's 
(i.e. the right panel of Fig.~\ref{fig:handbags}), 
the Quark Exchange Mechanism (QEM). 
Besides the QEM, $\rho^0$ electroproduction at large $Q^2$ and small
$x_B$ proceeds predominantly through a perturbative two-gluon exchange
mechanism (PTGEM) as studied in Ref.~\cite{Fra96}. 
To compare to the data at intermediate $Q^2$, the
power corrections due to the parton's intrinsic transverse momentum
dependence were implemented in both mechanisms 
(see Ref.~\cite{Vdh99} for details), which gives a
significant reduction at the lower $Q^2$. The results are
compared with the data in Fig.~\ref{fig:rhotot2}, showing that the PTGEM
explains well the fast increase of the cross section 
at high c.m. energy ($W$), 
but substantially underestimates the data at the lower energies. This is
where the QEM is expected to contribute since $x_B$ is then in the
valence region. The results including the QEM describe 
the change of behavior of the data at lower $W$ quite nicely. 

Recently, $\rho^0_L$ data have been obtained by the HERMES
Collaboration for $Q^2$ up to 5 GeV$^2$ and around 
$W \approx$ 5 GeV \cite{Air00}. These data show a clear dominance of the QEM 
in the intermediate $W$ range as predicted in \cite{Vdh98,Vdh99}. 
The model ansatz for the SPD's of Ref.~\cite{Vdh99} 
gives a fairly good agreement with these longitudinal
$\rho^0$ electroproduction data \cite{Air00}.

\subsection{Perspectives and outlook}

In order to extract SPD's from forthcoming data, 
the $Q^2$ evolution of the SPD's has already been worked
out at the next-to-leading order level \cite{Bel98}, which shows that  
the $Q^2$ evolution of the SPD's interpolates between the evolution of the
parton distributions and the evolution of the distribution amplitudes. 
Also radiative corrections to the coefficient functions 
have been calculated recently in next-to-leading order \cite{Bel99}.  
\newline
\indent
A major open theoretical question in this field is how the SPD's can be
deconvoluted from the leading order amplitudes. Suitable parametrizations
of the SPD's, incorporating all physical constraints, might be one
avenue to tackle this problem. 
In absence of a solution to this problem, one has to resort to model
calcuations or educated guesses for the SPD's in order to compare with
the data. 
\newline
\indent
On the experimental side, it is clear that new and accurate data are needed 
for various exclusive channels at large $Q^2$ in the valence region, where
the quark exchange mechanism dominates. 
Several experiments are being performed or are planned or 
proposed at JLab \cite{Guid98,Ber99,Ber00}, HERMES and 
COMPASS \cite{Poc99,d'Ho99}. 
Looking somewhat further into the future, the measurement of hard exclusive 
reactions will be one of the central themes for the planned upgrade of 
JLab to 12 GeV \cite{Bur00}.
A facility with high luminosity combined with an intermediate energy
of around 25 GeV, such as e.g. the ELFE project \cite{Bur99}, will be a
dedicated facility to measure these exclusive reactions at high momentum
transfer and to map out these new SPD's in detail.
Although such exclusive
experiments at large \( Q^{2} \) are quite demanding, 
the fundamental interest of the SPD's justifies 
an effort towards their experimental determination. 

\section{QED radiative corrections to virtual Compton scattering}
\label{radcorr}

\subsection{Introduction}

We have discussed in section~\ref{vcspol} how 
VCS below pion production threshold, allows us to access
generalized polarizabilities of the proton. 
Furthermore, we have seen in section~\ref{dvcs} 
that VCS in the Bjorken regime determines  
generalized parton distributions of the nucleon. 
In both regimes, experiments are either being 
done or planned for the near future. 
In order to extract the nucleon structure information of interest from 
the $e p \to e p \gamma$ reaction, especially in those
kinematical situations where the Bethe-Heitler process is not
negligible, it is indispensable to have a very good understanding 
of the QED radiative corrections to the $e p \to e p \gamma$ reaction. 
\newline
\indent
The $e p \rightarrow e p \gamma$ 
reaction is particular in comparison with other
electron scattering reactions, because the photon can be emitted from both
the proton side (this is the VCS process which contains the nucleon structure
information of interest) or from one of the electrons 
(which is the Bethe-Heitler process). 
The radiative corrections obtained from the Bethe-Heitler process 
differ formally from the case of elastic electron scattering.

\subsection{Results for the QED radiative corrections to VCS}

The first order QED radiative corrections to the $e p \to e p \gamma$
reaction were calculated in Ref.~\cite{Vdh00a}.
The one-loop virtual radiative
corrections have been evaluated by a combined analytical-numerical
method. Several tests were performed to cross-check the numerical method used. 
It was shown in Ref.~\cite{Vdh00a} 
how all IR divergences cancel when adding the
soft-photon emission processes. Furthermore, a 
fully numerical method was presented for the photon emission 
processes where the photon energy is not very small compared with the 
electron energies, which makes up the radiative tail.

\begin{figure}[h]
\vspace{-.5cm}
\begin{center}
\leavevmode
\hbox{
\epsfysize=10.0truecm
\epsfbox{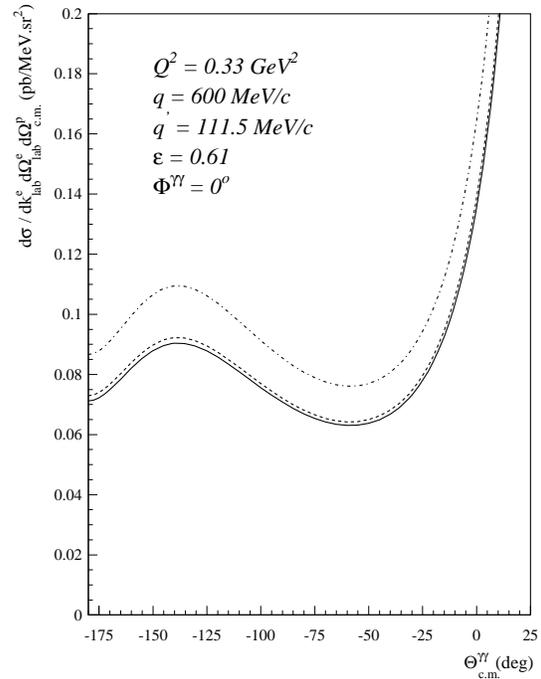} }
\end{center}
\centerline{\small \it  }
\vspace{-.8cm}
\caption{Differential $e p \to e p \gamma$ cross section for MAMI 
kinematics at ${\rm q}^{'}$ = 111.5 MeV/c. 
Dashed-dotted curve : BH + Born contribution, 
dashed curve : BH + Born + virtual radiative correction, 
full curve : BH + Born + total radiative correction. 
Figure from Ref.~\cite{Vdh00a}.} 
\label{fig:cross112_mami}
\end{figure}

Fig.~\ref{fig:cross112_mami} shows as representative result the
effect of the radiative corrections on the VCS differential cross
section for the MAMI VCS experiment \cite{Roc00} 
at an outgoing photon energy of ${\rm q}'$ = 111.5 MeV 
(we refer to \cite{Vdh00a} for all details and more results). 
It is seen that the total effect of the radiative corrections 
in these kinematics is a reduction of the BH+Born cross section 
of the order of 20\%.  
The effect of the radiative corrections 
was also confirmed by the experimental results at
very low energy of the outgoing photon (${\rm q}'$ = 33 MeV), 
where the effect of the GP's is negligible. 
From the difference between the radiatively corrected data 
and the BH + Born result, the two
values of Eq.~(\ref{eq:mamiexp}) for the combinations of the proton's 
GP's at $Q^2 \simeq$ 0.33 GeV$^2$ have been extracted in 
Ref.~\cite{Roc00}. 
\newline
\indent
In Ref.~\cite{Vdh00a}, calculations of the VCS radiative corrections were
also given for unpolarized and polarized VCS observables both at 
low energies and in the Bjorken regime. 
The results will be an important tool for the analysis of 
present and forthcoming VCS experiments, in order to extract the nucleon
structure information from these experiments.

\section{Conclusions and outlook}
\label{conclusion}

It has been discussed how the real and virtual Compton
scattering in different kinematical regimes provide 
new tools to extract nucleon structure information. 
\newline
\indent
It has been seen that for RCS at low energy, new accurate data
have become available which not only allow the extraction of scalar
polarizabilities of the proton, but also start to explore the spin
polarizabilities of the nucleon. Those spin polarizabilities have been
calculated recently to $O(p^4)$ in HBChPT. A fixed-t dispersion
relation formalism was developed to extract the nucleon
polarizabilities with a minimum of model dependence from both
unpolarized and polarized RCS data. The DR formalism was also used to
obtain information on new higher order polarizabilities of the proton,
providing new nucleon structure observables and a new testing ground
for the chiral calculations. 
\newline
\indent
The VCS reaction at low photon energy maps out
the spatial distribution of the polarization densities of the proton,
through generalized polarizabilities. Over the last few years, 
the VCS has become a mature field and a 
first experiment at MAMI at low energy has been successfully completed. 
In order to extract GP's from VCS data 
over a larger range of energies, a dispersion
relation formalism is underway, providing a new tool to analyze such
data. The DR formalism provides already results for 4 of the 6 GP's,
which can be confronted with chiral predictions. 
\newline
\indent
The RCS reaction at high energy and large momentum transfer is a tool
to access information on the partonic structure of the nucleon. 
The PQCD predictions for wide angle real Compton scattering 
(90$^o$ in the c.m.) show a strikingly different behavior 
than competing soft-overlap type mechanisms, and
forthcoming experiments can teach us about the interplay of both
mechanism at accessible energies.  
\newline
\indent
The VCS reaction in the Bjorken regime and associated hard
electroproduction reactions give access 
to new, generalized (skewed) parton distributions. 
The study of SPD's has opened up a whole new field in the study of 
nucleon structure. These observables unify two different fields
as they interpolate between purely inclusive quantities (parton
distributions) on the one hand and between simple exclusive quantities
(such as form factors) on the other hand. Besides the SPD's $H$ and $\tilde
H$, which reduce in the DIS limit to the quark distribution and quark
helicity distribution respectively, 
there are two entirely new leading twist
SPD's ($E$ and $\tilde E$), which cannot be accessed in DIS.   
The non-perturbative information contained in the SPD's is
rather rich as they are functions of 3 different variables. In particular, the
skewedness variable $\xi$ leads to different regions where one is
sensitive either to quark distribution type information or to meson
distribution amplitude information in the nucleon. 
Through a sum rule, two of these SPD's ($H$ and $E$) determine the
quark orbital angular momentum contribution to the nucleon spin. 
An educated guess was shown for these SPD's, which was used to 
estimate the leading order DVCS amplitude. 
Furthermore, the leading order meson
electroproduction amplitudes were discussed and compared to the available
data. In particular it was seen that in the intermediate $W$ range
(valence region), a dominance of the handbag mechanism is predicted
for $\rho^0$ electroproduction, which seems to be well confirmed by
recent HERMES data. 
Furthermore, the extension of the formalism of the SPD's 
to the $N \to \Delta$ transition was discussed. 
The large $N_c$ limit allows to relate the $N \to \Delta$
SPD's to the $N \to N$ SPD's.  
The transverse spin asymmetry was discussed as a promising observable,
likely to be less sensitive to higher twist effects.  
\newline
\indent
It is easy to foresee that the fields of real and virtual Compton scattering 
will show important activities in the near future both on the
theoretical and experimental sides, and that an attempt to review them is very
timely. It is hoped however, that the works initiated and discussed in this
paper will stimulate further efforts on the theoretical and
experimental sides to extend our knowledge of nucleon structure in new
directions.

\section*{Acknowledgements}

It is a pleasure for me to thank my collaborators 
who participated in an important way in the different works referred to in this
paper : N. D'Hose, D. Drechsel, L.L. Frankfurt, 
J.M. Friedrich, M. Gorchtein, P.A.M. Guichon, 
M. Guidal, B. Holstein, D. Lhuillier, D. Marchand, A. Metz, B. Pasquini, 
M.V. Polyakov, M. Strikman, L. Van Hoorebeke, and J. Van de Wiele.
I also like to thank the many experimental colleagues working in these
fields for numerous and very useful discussions. 
Furthermore, I would like to thank D. Drechsel 
also for a careful reading of the text. 
\newline
\indent
This work was supported by the Deutsche Forschungsgemeinschaft (SFB443).

\end{document}